\documentclass[11pt, oneside]{article}   	% use "amsart" instead of "article" for AMSLaTeX format
\usepackage{geometry}
\usepackage{xcolor}              		% See geometry.pdf to learn the layout options. There are lots.
\usepackage{amsmath}
\usepackage{amsfonts}
\usepackage{amssymb}
\usepackage{mathrsfs}
\usepackage[most]{tcolorbox}
\usepackage{array, bigdelim,multirow,multicol}
\usepackage{dsfont}
\usepackage{longtable}
\usepackage{ltablex}
\usepackage{enumitem}
\usepackage{cite}
\usepackage[colorlinks, linkcolor=red, anchorcolor=black, citecolor=blue]{hyperref}
\usepackage[framemethod=tikz]{mdframed}
\usepackage{environ}
\NewEnviron{myalign}{%
\begin{align}
\scalebox{0.9}{$\BODY$}
\end{align}
}
\usepackage{graphicx}				% Use pdf, png, jpg, or eps§ with pdflatex; use eps in DVI mode							% TeX will automatically convert eps --> pdf in pdflate
\usepackage{bbm}
\usepackage{bm}								
\newsavebox\myv
\newcommand{\ignore}[1]{}
\newcommand{\dd}{\displaystyle}

\newcommand{\nn}{\nonumber}
\newcommand{\be}{\begin{equation}}
\newcommand{\ee}{\end{equation}}
\newcommand{\bea}{\begin{eqnarray}}
\newcommand{\eea}{\end{eqnarray}}

\def\gapmatrix[{\begin{pmatrix}
\gaprows}
\def\gaprows#1[#2]#3{%
\gapcell#2\gapendrow,\ifx]#3\end{pmatrix}\else\afterfi\\\gaprows\fi}

\def\afterfi#1\fi{\fi#1}
\def\gapcell#1,{#1\uppercase{&}\gapcell}
\def\gapendrow#1\gapcell{}

\allowdisplaybreaks

\makeatletter
\renewcommand*{\@fnsymbol}[1]{\ensuremath{\ifcase#1\or *\or  \mathsection\or \ddagger\or
\dagger\or \mathparagraph\or \|\or **\or \dagger\dagger
\or \ddagger\ddagger \else\@ctrerr\fi}}
\makeatother

\begin{document}

\unitlength = 1mm

\setlength{\extrarowheight}{0.2 cm}

\title{
{\Large\bf
 Universal predictions of  Siegel modular invariant theories\\ near the fixed points
}
\\[0.cm]}
\date{}

\author{
Gui-Jun~Ding$^{1}$
\thanks{E-mail: {\tt dinggj@ustc.edu.cn}},
\
Ferruccio~Feruglio$^{2}$
\thanks{E-mail: {\tt feruglio@pd.infn.it}}
\ and
Xiang-Gan~Liu$^{3}$
\thanks{E-mail: {\tt xianggal@uci.edu }}
\
\\*[15pt]
\centerline{
\begin{minipage}{\linewidth}
\begin{center}
$^1${\small
Department of Modern Physics, University of Science and Technology of China,\\
Hefei, Anhui 230026, China}\\[2mm]
$^2${\small
INFN, Sezione di Padova, Via Marzolo~8, I-35131 Padua, Italy}\\[2mm]
$^3${\small
Department of Physics and Astronomy, University of California, Irvine, CA 92697-4575 USA}\\
\end{center}
\end{minipage}}
\\[8mm]}
\maketitle
\thispagestyle{empty}
\centerline{\large\bf Abstract}
\begin{quote}
\indent
We analyze a general class of locally supersymmetric, CP and modular invariant models of lepton masses depending on two complex moduli taking values in the vicinity of a fixed point, where the theory enjoys a residual symmetry under a finite group. Like in models that depend on a single modulus, we find that all physical quantities exhibit a universal scaling with the distance from the fixed point. There is no dependence on the level of the construction, the weights of matter multiplets and their representations, with the only restriction that electroweak lepton doublets transform as irreducible triplets of the finite modular group. Also the form of the kinetic terms, which here are assumed to be neither minimal nor flavor blind, is irrelevant to the outcome. The result is remarkably simple and the whole class of examined theories gives rise to five independent patterns of neutrino mass matrices. Only in one of them, the predicted scaling agrees with the observed neutrino mass ratios and lepton mixing angles, exactly as in single modulus theories living close to $\tau=i$.
\end{quote}

\newpage
%%%%%%%%%%%%%%%%%%%%%%%%%%%%%%%%%%%%%%%%%%%%%%%
\section{Introduction}
%%%%%%%%%%%%%%%%%%%%%%%%%%%%%%%%%%%%%%%%%%%%%%%
The flavour puzzle is one of the big unsolved problems in particle physics. The observed regularities in the fermion mass spectrum and the mixing angles still escape our understanding and their consistent description requires more than twenty free parameters. One of the tools that have been exploited so far to tackle the puzzle relies on flavour symmetries, acting in generation space and supposed to reduce the number of flavour parameters, including those associated with CP violation. In a bottom-up approach, the freedom of the model builder is however disarmingly huge. Not only we can freely choose flavour groups and representations of matter fields but, more importantly,
we should add an ad hoc symmetry-breaking sector to reconcile the approximate predictions of the exact symmetry limit with the accurate experimental data. The complicated architecture of such a sector often results in a limited predictive power of the whole approach.

Modular invariance offers in principle some advantage over traditional flavour symmetries of the above type~\cite{Feruglio:2017spp}. The starting point is the symmetry-breaking sector itself consisting, in the simplest case, of a single complex field, the modulus, living in the upper half of the complex plane and transforming nontrivially under the modular group.
In a supersymmetric realization~\cite{Ferrara:1989bc,Ferrara:1989qb}, modular invariance severely restricts the Yukawa couplings, which are required to be modular forms, holomorphic functions of the modulus with appropriate transformation properties under the modular group. Radiative corrections and supersymmetry breaking effects are negligible in a large portion of the parameter space~\cite{Criado:2018thu}.
Invariance under the modular group offers a simple, axion-free solution to the strong CP problem~\cite{Feruglio:2023uof}.
Moreover, such a framework is intimately related to the basic properties of superstring compactifications~\cite{Hamidi:1986vh,Dixon:1986qv,Lauer:1989ax,Lauer:1990tm,Erler:1991nr,Stieberger:1992vb,Erler:1992gt,Stieberger:1992bj,Giveon:1994fu,Cremades:2003qj,Blumenhagen:2005mu,Abel:2006yk,Blumenhagen:2006ci,Marchesano:2007de,Antoniadis:2009bg,Kobayashi:2016ovu,Kobayashi:2020hoc,Cremades:2004wa,Abe:2009vi,Kikuchi:2020frp,Kikuchi:2020nxn,Ramos-Sanchez:2024keh},
allowing the bottom-up and top-down approaches to reinforce each other~\cite{Nilles:2023shk}.
An intense activity in both directions has been pursued in the last years~\cite{Kobayashi:2023zzc,Ding:2023htn}.

These nice features are partially spoiled by the freedom associated with the transformation law of the matter fields under the modular group and with the choice of kinetic terms. The former requires the specification of a level $n$ and, for each matter multiplet $\varphi^{(I)}$, a weight $k_I$ and a representation $\rho_I$ of the adopted finite modular group. The latter affects the fermion mass spectrum~\cite{Chen:2019ewa}, except for the case of minimal or flavor universal K\"ahler potential. Some progress in dealing with this problem has been made with the introduction of eclectic flavour symmetries~\cite{Baur:2019kwi,Nilles:2020nnc,Nilles:2020kgo,Nilles:2020tdp,Baur:2020jwc,Nilles:2020gvu,Baur:2021mtl},
but the prize is the reintroduction of a nonminimal symmetry-breaking sector~\cite{Chen:2021prl,Baur:2022hma,Ding:2023ynd,Li:2023dvm}.
Even considering the lepton sector alone, by exploiting the existing freedom a large number of models correctly reproducing neutrino masses,
lepton mixing angles, and predicting leptonic CP-violating phases have been formulated~\cite{Kobayashi:2023zzc,Ding:2023htn}. The variety of different realizations allowed in a pure bottom-up approach has not allowed so far to identify a unique scenario.

Despite our inability to designate a single successful theory, from all the existent effective models we might have learned something significant about the principle underlying the fermion sector. Indeed, the leptonic models formulated so far suggest a preference for a value of the modulus near the self-dual fixed point $\tau_0=i$~\cite{Feruglio:2022koo}, which preserves the symmetry under $\tau\to-1/\tau$. Such a preference is even more pronounced for the subset of CP-invariant models, where the violation of CP is spontaneous and entirely controlled by the modulus $\tau$. Indeed, if electroweak lepton doublets are assigned to irreducible triplets of the finite modular group, a typical choice allowing to minimize the number of free parameters, the behavior of modular invariant models of lepton masses near the fixed points $\tau_0=i$ and $\tau_0=-1/2+i~\sqrt{3}/2$ is universal~\cite{Feruglio:2023mii}. All physical quantities scale with the distance $|\tau-\tau_0|$ in a way that is largely independent of the level $n$, the weight $k_I$, the specific representations $\rho_I$ and even the form of kinetic terms. Only a few patterns of neutrino mass matrices can be realized, depending on the chosen fixed point. In particular, near $\tau_0=i$
almost all the successful models predict a normal ordering of neutrino masses, with mixing angles and neutrino mass differences that are all accommodated by $|\tau-\tau_0|\approx 0.1$. The key feature under this universal behavior is the residual $\mathbbm{Z}_4$ symmetry enjoyed by the theory at the fixed point, spontaneously broken by $(\tau-\tau_0)$.
In the bottom-up approach, the modulus $\tau$ is treated as a free parameter, optimized to maximize the agreement with the data. Remarkably, fixed points $\tau_0$ are extrema of a modular invariant energy density~\cite{Font:1990nt,Cvetic:1991qm}. Moreover, minima of the energy density close to but distinct from the fixed points have been established in modular invariant theories~\cite{Novichkov:2022wvg,Leedom:2022zdm,Ishiguro:2020tmo,Ishiguro:2022pde,Knapp-Perez:2023nty}.
Cosmological evolution can offer a mechanism for moduli trapping near the points enjoying an enhanced symmetry
\cite{Kofman:2004yc,Enomoto:2013mla,Kikuchi:2023uqo}.
In general, $(\tau-\tau_0)$ provides a useful expansion parameter to understand the hierarchy among charged
fermion masses and mixing angles~\cite{Feruglio:2021dte,Novichkov:2021evw,Petcov:2022fjf,Kikuchi:2023cap,Abe:2023ilq,Kikuchi:2023jap,Abe:2023qmr,Petcov:2023vws,Abe:2023dvr,deMedeirosVarzielas:2023crv,Petcov:2023fwh,Ishiguro:2024xph}.

Theories with a single modulus might offer an oversimplified description of the fermion mass spectrum. For example, in superstring compactifications fermion masses typically depend on several moduli. It is thus natural to ask whether a universal behavior persists in multi-moduli theories in the vicinity of the fixed points, where a residual symmetry under a discrete group $G_0$ applies. Moreover, if such universality is indeed exhibited in these richer theories, there is the chance of finding
new realistic patterns for fermion mass matrices. An attractive generalization of the upper half of the complex plane and the modular group $SL(2,\mathbbm{Z})$ is the Siegel upper half plane, where the Siegel modular group $Sp(2g,\mathbbm{Z})$ operates. The Siegel upper half plane has complex dimension $g(g+1)/2$ and coincides with the ordinary upper half complex plane
when the genus $g$ is one. Higher genera are realized in string theory compactification~\cite{Mayr:1995rx,Stieberger:1998yi,Ishiguro:2020nuf,Baur:2020yjl,Ishiguro:2021ccl}.
Bottom-up realizations have been formulated in ref.~\cite{Ding:2020zxw}, and their spontaneous CP breaking has been analyzed in ref.~\cite{Ding:2021iqp}.

The purpose of the present paper is to study the simplest possible such generalization, at genus $g=2$, and its property in the vicinity of the fixed points, which have been fully classified in~\cite{gottschling1961fixpunkte,gottschling1961fixpunktuntergruppen,gottschling1967uniformisierbarkeit}.
We focus on the lepton sector, adopting completely generic level $n$,
weights $k_I$ of the involved multiplets and representations $\rho_I$,
with the only assumption that the electroweak lepton doublets $L$ are assigned to an irreducible triplet of the relevant modular group. We also allow the kinetic terms to be the most general ones. These features are implemented in a CP-invariant locally supersymmetric theory.
There are strong indications that the four-dimensional CP symmetry is a gauge symmetry
~\cite{Dine:1992ya,Choi:1992xp,Leigh:1993ae}, even starting from a higher-dimensional theory where CP is not conserved.
It has been conjectured, as a general property of string theory, that CP is indeed a gauge symmetry of the four-dimensional theory. In this context, CP can only be violated spontaneously, by complex expectation values of fields, in our case the moduli.
The moduli will be restricted to a suitable region of the Siegel upper half plane, invariant under a subgroup of $Sp(4,\mathbbm{Z})$ hosting three-dimensional irreducible representations of the finite modular group. The widest such a region has complex dimension two and provides a nontrivial extension of the framework studied in ref.~\cite{Feruglio:2022koo,Feruglio:2023mii}.

In Section 2 we will review the Siegel modular group $Sp(4,\mathbbm{Z})$ and the two-dimensional invariant regions of the Siegel upper half plane. We identify the region that allows to assign lepton doublets to irreducible triplets of the relevant finite modular group and we analyze the fixed points belonging to this region. In Section 3 we define our CP-invariant locally supersymmetric theory and we describe the requirements for modular invariance. At the fixed points the theory enjoys a residual symmetry under a finite group $G_0$. An important step is provided by a field redefinition that, linearizing the action of $G_0$, considerably simplifies our task.
In Section 4, for each fixed point we identify the group $G_0$ and we provide the decomposition of all irreducible triplets of the finite modular group under $G_0$. This information is sufficient to find the pattern of neutrino mass matrices, in the basis where kinetic terms are canonical and charged lepton mass matrices diagonal, expressed as a series expansion in powers of $(\tau-\tau_0)$. Details on the group theory of the residual symmetry $G_0$ can be found in the Appendices.

The result is remarkably simple. Apart from the unrealistic case of a neutrino mass matrix vanishing to all orders of the $(\tau-\tau_0)$ expansion, only five patterns are found. Four of them coincide with those arising in $SL(2,\mathbbm{Z})$-invariant single modulus theories in the vicinity of the fixed points $\tau_0=i$ and $\tau_0=-1/2+i~\sqrt{3}/2$. In particular, one of these four patterns is especially effective in accommodating the existing data, with no required tuning of the free parameters. The last pattern predicts neutrino masses of the same order of magnitude and mixing angles of approximately the same size and does not provide any explanation for the smallness of $\Delta m^2_{sol}/\Delta m^2_{atm}$ and $\sin^2\theta_{13}$. It is intriguing that, not only the universal behavior of the theory near the fixed points is confirmed, but essentially no realistic patterns of neutrino mass matrices different from those found in the single modulus case are exhibited in this class of multi-moduli theories.

%%%%%%%%%%%%%%%%%%%%%%%%%%%%%%%%%%%%%%%%%%%%%%%
\section{Siegel modular group}
%%%%%%%%%%%%%%%%%%%%%%%%%%%%%%%%%%%%%%%%%%%%%%%

One of the most natural generalization of theories invariant under the modular group $SL(2,\mathbbm{Z})$, where masses and mixing angles depend on a single complex modulus $\tau$, is the class of theories invariant under the Siegel (or symplectic) modular group $Sp(4,\mathbbm{Z})$~\cite{Ding:2020zxw}, where moduli are described by a two-by-two complex symmetric matrix $\tau$ belonging to the Siegel upper-half plane ${\cal H}_2$:
\be
\label{modulus}
\tau=\left(
\begin{array}{cc}
\tau_1&\tau_3\\
\tau_3&\tau_2
\end{array}
\right)\,,~~~~~\det({\tt Im}(\tau))>0\,,~~~~~{\rm Tr}({\tt Im}(\tau))>0\,.
\ee
Assuming CP and ${\cal N}=1$ supersymmetric invariance, and neglecting gauge interactions, these theories involve a set of chiral supermultiplets, $\Phi=(\tau,\varphi^{(I)})$, where $\tau$ is dimensionless and gauge-invariant. The Siegel modular group $Sp(4,\mathbbm{Z})$, adopted as flavour symmetry, acts on $\Phi=(\tau,\varphi^{(I)})$ as~\cite{Ding:2020zxw}:
\be
\left\{
\begin{array}{l}
\tau\to \gamma \tau=(A \tau+B)(C\tau +D)^{-1}\,,
\\[0.2 cm]
\varphi^{(I)}\to [\det(C\tau+D)]^{-k_I} \rho_I(\gamma) \varphi^{(I)}~\,,
\end{array}
\right.~~~~~~~\gamma=\begin{pmatrix}
A & B \\
C & D
\end{pmatrix}\in Sp(4,\mathbbm{Z})\,.
\label{eq:tmg-1}
\ee
with suitable conditions on the submatrices $A$, $B$, $C$ and $D$. The weight $k_I$ is assumed to be an integer and $\rho_I(\gamma)$ denotes a unitary representation of a finite copy $\Gamma_n$ of $Sp(4,\mathbbm{Z})$~\footnote{For a generic genus $g$, the Siegel modular group is $Sp(2g,\mathbbm{Z})$ and the finite modular groups are denoted by $\Gamma_{g,n}$. Here we use the concise notation $\Gamma_n=\Gamma_{2,n}$.}.

Such a finite copy, known as finite Siegel modular group, is defined as the quotient group $\Gamma_n=Sp(4,\mathbbm{Z})/\Gamma(n)$ where $\Gamma(n)$
is the principal congruence subgroup $\Gamma(n)$ of level $n$:
\begin{equation}
\label{eq:def_principal congruence subgroup}
\Gamma(n)=\Big\{\gamma \in Sp(4,\mathbbm{Z}) ~\Big|~  \gamma \equiv \mathbbm{1} \,\texttt{mod}\, n\Big\}\,,
\end{equation}
$n$ being a generic positive integer.
The finite Siegel modular group has finite order~\cite{Koecher1954Zur,thesis_Fiorentino}:
\begin{eqnarray}
|\Gamma_n|=n^{10}\prod_{p|n}(1-\dfrac{1}{p^{2}})(1-\dfrac{1}{p^{4}})\,,
\end{eqnarray}
where the product is over the prime divisors $p$ of $n$. $|\Gamma_n|$ rapidly grows with $n$: for example $|\Gamma_2|=720$, $|\Gamma_3|=51840$. The group $\Gamma_2$ is isomorphic to $S_6$, and $\Gamma_3$ is $Sp(4, F_3)$, the double covering of Burkhardt group. These finite groups do not possess three-dimensional irreducible representations
suitable to accommodate three fermion generations.
The smallest irreducible representation with degree bigger than one is 5 for $\Gamma_2$ and 4 for $\Gamma_3$. If $n$ is a power of an odd prime, the next lowest dimensional (complex, irreducible) representations of $\Gamma_n$, after the trivial representation, are two of dimension $(n^2-1)/2$ and two of dimension $(n^2+1)/2$~\cite{katz2020rigid}. In this series we do not find
any three-dimensional irreducible representation. It is quite possible that also when $n$ is a power of an even prime, the smallest nontrivial irreducible representation of $\Gamma_n$ are quite large. Though we do not have a formal proof, we suspect that none of the finite Siegel modular groups $\Gamma_n$ possesses three-dimensional irreducible representations. The requirement of three-dimensional irreducible representations is not mandatory, but is a very convenient one since it reduces the number of independent parameters needed to describe the fermion mass spectrum of the theory.
%%%%%%%%%%%%%%%%%%%%%%%%%%%%%%%%%%%%%%%%%%%%%%%%%
\subsection{Invariant regions}
%%%%%%%%%%%%%%%%%%%%%%%%%%%%%%%%%%%%%%%%%%%%%%%%%
In ref.~\cite{Ding:2020zxw} we have shown how to overcome this difficulty, by restricting $\tau$ to a convenient region $\Sigma$ of the Siegel upper-half plane ${\cal H}_2$, invariant under a subgroup of $Sp(4,\mathbbm{Z})$.
Individual points in $\Sigma$ are left invariant by a common subgroup $H$ of $Sp(4,\mathbbm{Z})$, the stabilizer
\be
H~\tau=\tau\,.
\label{fp}
\ee
The region $\Sigma$, as a whole, is left invariant by the normalizer
\be
N(H)=\left\{\gamma\in Sp(4,\mathbbm{Z})\vert \gamma^{-1}H\gamma=H \right\}\,.
\ee
In general, $H$ is a proper subgroup of $N(H)$. If $\Sigma$ consists of a single point, the stabilizer and the normalizer coincide. In our theory, we can consistently restrict the domain of moduli to a region $\Sigma$ of this type, and replace $Sp(4,\mathbbm{Z})$ with $N(H)$ and $\Gamma_n$ with $N_n(H)$, where the integer $n$ is the level of the representation and $N_n(H)$ is a finite copy of the normalizer, obtained through the same steps leading to the Siegel finite modular groups $\Gamma_n$. The group $N(H)$ acts on $\Phi=(\tau,\varphi^{(I)})$ as in eq.~\eqref{eq:tmg-1}, where $\gamma$ belongs to $N(H)$ and $\rho_I(\gamma)$ is a unitary representation of $N_n(H)$. Since in general $N_n(H)$ is smaller than $\Gamma_n$, there is a chance that it possesses three-dimensional irreducible representations.

Finally, up to Sigel modular transformations, CP transformations on the chiral multiplets read (we use a bar to denote conjugation of fields)~\cite{Baur:2019kwi,Novichkov:2019sqv,Baur:2019iai,Ishiguro:2020nuf,Baur:2020yjl,Ding:2021iqp}:
\begin{align}
\label{CP}
\tau\xrightarrow{\mathcal{CP}}-\bar\tau\,,~~~~~~
\varphi^{(I)} \xrightarrow{\mathcal{CP}} X_I\;\bar\varphi^{(I)}\,,
\end{align}
where the CP transformation matrix $X_I$ is fixed by the following consistency conditions~\cite{Ding:2021iqp}
\begin{equation}
\label{eq:gCP-cons-cond}X_I\,\rho^{*}_I(S)X^{-1}_I=\rho_I(S^{-1}),~~~~X_I\,\rho^{*}_I(T_i)X^{-1}_I=\rho_I(T_i^{-1})\,,
\end{equation}
up to an overall irrelevant phase. Here $S$ and $T_i\; (i=1, 2, 3)$ denote the generators of $Sp(4,\mathbbm{Z})$:
\begin{equation}
\label{eq:Siegel-generators}
S=\begin{pmatrix}
0 & \mathbbm{1}_2\\
-\mathbbm{1}_2 & 0
\end{pmatrix}\,,~~~~T_i=\begin{pmatrix}
\mathbbm{1}_2 & B_i\\
0 & \mathbbm{1}_2
\end{pmatrix}
\end{equation}
with
\begin{equation}
B_1 =\left(\begin{array}{cc}
1 & 0 \\
0 & 0 \\
\end{array}
\right)\,,~~~ B_2=\left(\begin{array}{cc}
0 & 0 \\
0 & 1 \\
\end{array}
\right)\,, ~~~ B_3=\left(\begin{array}{cc}
0 & 1 \\
1 & 0 \\
\end{array}
\right)\,.
\end{equation}
In the basis where the unitary representation matrices $\rho_I(S)$ and $\rho_I(T_i)$ are symmetric, the consistency conditions of eq.~\eqref{eq:gCP-cons-cond} is solved by $X_I=\mathbbm{1}$ which is the canonical CP transformation.
In our analysis, it will be convenient to make use of generalized CP transformations, which combine eq. (\ref{CP}) with a modular transformation. They will be discussed in Section \ref{afr}.
%%%%%%%%%%%%%%%%%%%%%%%%%%%%%%%%%%%%%%%%%%%%%%%%%%%%%%%%%
\subsection{Two-dimensional invariant regions}
%%%%%%%%%%%%%%%%%%%%%%%%%%%%%%%%%%%%%%%%%%%%%%%%%%%%%%%%%
Having to abandon the full Siegel upper-half plane ${\cal H}_2$, the widest regions $\Sigma$ have complex dimension two.
In the Siegel upper-half plane ${\cal H}_2$, there are two such regions, left invariant by the action of
a subgroup $N(H)$ of $Sp(4,\mathbbm{Z})$~\cite{gottschling1961fixpunkte,gottschling1961fixpunktuntergruppen,gottschling1967uniformisierbarkeit}.
\begin{table}[h!]
\centering
%\resizebox{0.65\textwidth}{!}{
\begin{tabular}{|c|c|c|} \hline\hline
Invariant region $\tau$ &  Stabilizer $H$ & $\begin{array}{l}{\rm generators~ of~ the}\\ {\rm Normalizer}~ N(H)\end{array}$  \\  \hline\hline
 $\Sigma_1=\begin{pmatrix} \tau_1 & 0 \\ 0 & \tau_2 \end{pmatrix}$
 &$Z_2\times Z_2$& eq.~\eqref{eq:NorTwo1} \\ \hline
 $\Sigma_2=\begin{pmatrix} \tau_1 & \tau_3 \\ \tau_3 & \tau_1 \end{pmatrix}$ & $Z_2\times Z_2$&eq.~\eqref{eq:NorTwo2}   \\ \hline\hline
 \end{tabular}
%}
\caption{\label{IR2}
Invariant regions $\Sigma_{1,2}$ of complex dimension two in the Siegel upper half plane $\mathcal{H}_2$. The generators of the corresponding normalizers are shown in eqs.~\eqref{eq:NorTwo1} and \eqref{eq:NorTwo2}.}
\end{table}
%%%%%%%%%%%%%%%%%%%%%%%%%%%%%%%%%%%%%%%%%%%%%%%%%%%%%%%%%
\subsubsection{\bf $\Sigma_1$}
%%%%%%%%%%%%%%%%%%%%%%%%%%%%%%%%%%%%%%%%%%%%%%%%%%%%%%%%%
\noindent
One of them is
\be
\Sigma_1=\left\{\begin{pmatrix} \tau_1 & 0 \\ 0 & \tau_2 \end{pmatrix}\in {\cal H}_2\right\}\,.
\ee
The stabilizer $H$ is the $\mathbbm{Z}_2\times \mathbbm{Z}_2$ group whose generators are:
\begin{myalign}
\label{StaTwo1}
-\mathbbm{1}=\begin{pmatrix} -1&0&0&0 \\ 0&-1&0&0 \\ 0&0&-1&0 \\ 0&0&0&-1\end{pmatrix},~~~~~~~~~h=\begin{pmatrix} 1&0&0&0 \\ 0&-1&0&0 \\ 0&0&1&0 \\ 0&0&0&-1\end{pmatrix}\,.
\end{myalign}
The normalizer $N(H)$ is generated by the elements:
\begin{align}
&G_1=\begin{pmatrix} 1&0&0&0 \\ 0&0&0&1 \\ 0&0&1&0 \\ 0&-1&0&0\end{pmatrix}\,,~~~~
G_2=\begin{pmatrix} 1&0&0&0\\ 0&1&0&1 \\ 0&0&1&0 \\ 0&0&0&1 \end{pmatrix}\,,~~~~
G'_1=\begin{pmatrix} 0&0&1&0\\ 0&1&0&0 \\ -1&0&0&0 \\ 0&0&0&1 \end{pmatrix},\nn\\
\label{eq:NorTwo1}& G'_2=\begin{pmatrix}
1&0&1&0\\ 0&1&0&0 \\ 0&0&1&0 \\ 0&0&0&1 \end{pmatrix}\,,~~~~
G_3=\begin{pmatrix} 0&1&0&0\\ 1&0&0&0 \\ 0&0&0&1 \\ 0&0&1&0 \end{pmatrix}\,.
\end{align}
The action of the group generators on the moduli $\tau_1$ and $\tau_2$ is
\begin{eqnarray}
\nonumber&& G_1: \tau_1\rightarrow\tau_1,\qquad  \tau_2\rightarrow-\frac{1}{\tau_2}\,,\\
\nonumber&& G_2: \tau_1\rightarrow\tau_1,\qquad  \tau_2\rightarrow\tau_2+1\,,\\
\nonumber&& G'_1: \tau_1\rightarrow-\frac{1}{\tau_1},\qquad  \tau_2\rightarrow\tau_2\,,\\
\nonumber&& G'_2: \tau_1\rightarrow\tau_1+1,\qquad  \tau_2\rightarrow\tau_2\,,\\
&& G_3: \tau_1\rightarrow\tau_2\qquad  \tau_2\rightarrow\tau_1.
\end{eqnarray}
The generators $G_{1,2}$, $G'_{1,2}$ and $G_3$ obey the following relations
\begin{eqnarray}
\nonumber&G^4_1=(G_1G_2)^3=1,~~~G^2_1G_2=G_2G^2_1,~~~G'^4_1=(G'_1G'_2)^3=1,~~~G'^2_1G'_2=G'_2G'^2_1\,,\\
\nonumber& G_1G'_1=G'_1G_1,~ G_1G'_2=G'_2G_1,~ G_2G'_1=G'_1G_2,~ G_2G'_2=G'_2G_2\,,\\
%\label{eq:Sigma1-gen}&G^2_3=1,~~~G_3G_1G^{-1}_3=G'_1,~~~G_3G_2G^{-1}_3=G'_2\,.
\label{eq:Sigma1-gen}&G^2_3=1,~~~G_3G_1=G'_1G_3,~~~G_3G_2=G'_2G_3\,.
\end{eqnarray}
Therefore the normalizer $N(H)$ is isomorphic to $\left(SL(2,\mathbb{Z})\times SL(2,\mathbb{Z})\right) \rtimes (\mathbbm{Z}_2)_M$, where the last factor, generated by $G_3$ that exchanges $\tau_1$ and $\tau_2$, represents
the so-called mirror symmetry in the string theory context. We find that the normalizer $N_n(H)$ corresponding to region
$\Sigma_1$  has no three-dimensional irreducible representations.
The triplet representations of the group $SL(2,\mathbb{Z}_n)\times SL(2,\mathbb{Z}_n)$ can be obtained from the direct product of $SL(2,\mathbb{Z}_n)$ singlets with the irreducible triplets of another $SL(2,\mathbb{Z}_n)$. The two triplet representations $SL(2,\mathbb{Z}_n)\times SL(2,\mathbb{Z}_n)$ related by the mirror symmetry $(\mathbbm{Z}_2)_M$ would form a six dimensional representation of the normalizer $N_n(H)$. Hence $N_n(H)$ has no three-dimensional irreducible representations. We tested this argument by varying $n$ from 2 to 18, and indeed we found no three-dimensional irreducible representations for $N_n(H)$, but only irreducible representations of
one, two, four, six, eight, nine, and higher dimension. Thus the region $\Sigma_1$ is only suitable to describe
electroweak lepton doublets transforming in a reducible representation of the finite normalizer. We do not consider
such a case here and we proceed by examining the region $\Sigma_2$.
%%%%%%%%%%%%%%%%%%%%%%%%%%%%%%%%%%%%%%%%%%%%%%%%%%%%%%%%%
\subsubsection{\bf $\Sigma_2$}
\label{S2}
%%%%%%%%%%%%%%%%%%%%%%%%%%%%%%%%%%%%%%%%%%%%%%%%%%%%%%%%%
\noindent
The other two-dimensional region is
\be
\Sigma_2=\left\{\begin{pmatrix} \tau_1 & \tau_3 \\ \tau_3 & \tau_1 \end{pmatrix}\in {\cal H}_2\right\}\,.
\ee
The stabilizer $H$ is the $\mathbbm{Z}_2\times \mathbbm{Z}_2$ group generated by:
\begin{myalign}
\label{StaTwo2}
-\mathbbm{1}=\begin{pmatrix} -1&0&0&0 \\ 0&-1&0&0 \\ 0&0&-1&0 \\ 0&0&0&-1\end{pmatrix},~~~~~~~~~~
h=\begin{pmatrix} 0&1&0&0 \\ 1&0&0&0 \\ 0&0&0&1 \\ 0&0&1&0\end{pmatrix}.
\end{myalign}
The normalizer $N(H)$ is generated by:
\begin{align}
&G_1=\begin{pmatrix} 1&0&1&0 \\ 0&1&0&1 \\ 0&0&1&0 \\ 0&0&0&1\end{pmatrix},~~~~~
G_2=\begin{pmatrix} 1&0&0&1\\ 0&1&1&0 \\ 0&0&1&0 \\ 0&0&0&1 \end{pmatrix},\nn \\
\label{eq:NorTwo2}
& G_3=\begin{pmatrix} 0&0&1&0\\ 0&0&0&1 \\ -1&0&0&0 \\ 0&-1&0&0 \end{pmatrix},~~~~~
G_4=\begin{pmatrix} 1&0&0&0 \\ 0&-1&0&0 \\ 0&0&1&0 \\ 0&0&0&-1\end{pmatrix}\,.
\end{align}
The action of the group generators on the moduli $\tau_1$ and $\tau_2$ is
\begin{eqnarray}
\nonumber&& G_1: \tau_1\rightarrow\tau_1+1,\qquad  \tau_3\rightarrow\tau_3\,,\\
\nonumber&& G_2: \tau_1\rightarrow\tau_1,\qquad  \tau_3\rightarrow\tau_3+1\,,\\
\nonumber&& G_3: \tau_1\rightarrow\frac{\tau_1}{\tau^2_3-\tau^2_1},\qquad  \tau_3\rightarrow-\frac{\tau_3}{\tau^2_3-\tau^2_1}\,,\\
&& G_4: \tau_1\rightarrow\tau_1,\qquad  \tau_3\rightarrow-\tau_3\,.
\end{eqnarray}
The generators $G_{1}$, $G_{2}$, $G_{3}$ and $G_4$ fulfill the following relations
\begin{eqnarray}
\nonumber& G^2_3=R,~~~R^2=G^2_4=(G_1G_3)^3=(G_2G_3)^6=(G_2G_4)^2=1\,,\\
\nonumber&G_2G_3G^2_2G_3G_2=G^3_3G^{-2}_2G_3,~~~G_1G_2=G_2G_1,~~~G_1G_4=G_4G_1,~~~G_3G_4=G_4G_3\,,\\
\label{eq:Sigma2-gen}& G_1R=RG_1,~~~G_2R=RG_2,~~~G_3R=RG_3,~~~G_4R=RG_4\,,
\end{eqnarray}
where $R=-\mathbbm{1}$.
\vskip 0.2 cm
\noindent
In table~\ref{irretrip}, for $n$ small and equal to a power of two or an odd prime, we show the number of irreducible triplets of $N_n(H)$. We see that when the level $n$ is a prime (or a power of primes) different from two, there are no three-dimensional irreducible representations, at least for the first few values of $n$. When $n$ is a power of two, there are several
three-dimensional irreducible representations and their number grows with $n$, until $n$ is equal to 8. We find that for $n=2,4,8$ there are $4,24,56$ three-dimensional irreducible representations, and for $n=16$ there are still only 56 three-dimensional irreducible representations. Moreover, there are 24 independent singlet representations in all $N_n(H)$.
\begin{table}[h!]
\begin{center}
\begin{tabular}{|c|l|c|c|}
\hline\hline
level $n$& {\tt order}& {\tt n. of conjugacy classes}& {\tt n. of irreducible triplets}\\
\hline
2&48&10&4\\
\hline
4&{6144}&{~109}&{~24}\\
\hline
8&{393216}&{~938}&{~56}\\
\hline
16&{25165824}&{~5572}&{~56}\\
\hline
32&{1610612736}&{~?}&{~?}\\
\hline
...&...&...&...\\
\hline
3&1152&35&0\\
\hline
5&28800&54&0\\
\hline
7&225792&77&0\\
\hline
11&3484800&135&0\\
\hline
13&9539712&170&0\\
\hline
...&...&...&...\\
\hline\hline
\end{tabular}
\caption{Orders, number of conjugacy classes and of irreducible triplets of the finite normalizers $N_n(H)$ in the region $\Sigma_2$. The results have been obtained with the software GAP.}
\label{irretrip}
\end{center}
\end{table}
We conjecture that there are no new three-dimensional irreducible representations for $n>8$. Moreover, the 56 three-dimensional irreducible representations at level 8 include those al the lower levels $n<8$. If our conjecture is correct, all irreducible triplets of $N_n(H)$ can be obtained from the product of the 24 singlets and 56 triplets of $\Gamma_{2,4,8}$.
Out of the resulting 1344 representations, we find that the inequivalent ones are 168. In the rest of this paper, we assume that these 168 representations exhaust the number of inequivalent irreducible triplets.
%%%%%%%%%%%%%%%%%%%%%%%%%%%%%%%%%%%%%%%%%%%%%%%%%
\subsection{Fixed points}
%%%%%%%%%%%%%%%%%%%%%%%%%%%%%%%%%%%%%%%%%%%%%%%%%
We analyze a class of theories, where the matrix $\tau$ is restricted to $\Sigma_{2}$ and the flavor group is the corresponding normalizer $N(H)$ defined in eq.~\eqref{eq:NorTwo2}. In a generic point of $\Sigma_{2}$, the flavour symmetry is broken down to the stabilizer $H$, eq.~\eqref{StaTwo2}. An exception is provided by the fixed points $\tau_0$ of $\Sigma_{2}$,
where the residual symmetry group $G_0$, a subgroup of $N(H)$, is bigger than $H$. The inequivalent fixed points of $Sp(4,\mathbbm{Z})$ in the upper half-plane are displayed in table~\ref{FixedPoints}. Fixed points 2 and 5 belong to $\Sigma_2$. Fixed points 3 and 4 belong to both $\Sigma_1$ and $\Sigma_2$. Here we will only consider the latter option. The fixed point 6 belongs to the region $\Sigma_1$ and will not be discussed. Finally, the fixed point 1, where CP is preserved, does not belong to either $\Sigma_1$ or $\Sigma_2$. We have also checked that it is not related by any $Sp(4,\mathbbm{Z})$ transformation to one point in either $\Sigma_1$ or $\Sigma_2$. This point will be dismissed in our analysis, too. 
\begin{table}[h!]
\centering
\resizebox{1.0\textwidth}{!}{
\begin{tabular}{|c|c|c|c|c|} \hline\hline
&Fixed points $\tau$ &  Residual symmetry in $Sp(4,\mathbbm{Z})$ &  $G_0=$~Residual symmetry in $N(H)$ &CP   \\  \hline\hline
1.& $\begin{pmatrix} \zeta & \zeta+\zeta^{-2} \\ \zeta+\zeta^{-2} & -\zeta^{-1} \end{pmatrix} $  &$Z_{10}$ & ---& $+$\\ \hline
2.& $\begin{pmatrix} \eta & \frac{1}{2}(\eta -1) \\ \frac{1}{2}(\eta -1) & \eta \end{pmatrix} $& $GL(2,3)$   &$D_4$ ~~$(\tau\in \Sigma_2$)& $+$\\ \hline
3.& $\begin{pmatrix} i & 0 \\ 0 & i \end{pmatrix}$ & $(Z_4\times Z_4)\rtimes Z_2$  &
$\left\{\begin{array}{cc}(Z_4\times Z_4) \rtimes Z_2&(\tau\in\Sigma_1)\\D_4\circ Z_4&(\tau\in\Sigma_2)\end{array}\right.$
&$+$ \\ \hline
4.& $\begin{pmatrix} \omega & 0 \\ 0 & \omega \end{pmatrix}$
 & $[72,30]\cong (Z_6\times Z_6)\rtimes Z_2$ &$\left\{\begin{array}{cc}(Z_6\times Z_6)\rtimes Z_2&(\tau\in\Sigma_1)\\ D_4\times Z_3 & (\tau\in\Sigma_2)\end{array}\right.$& $+$ \\ \hline
5.& $\dfrac{i\sqrt{3}}{3}\begin{pmatrix} 2 & 1 \\ 1 & 2 \end{pmatrix}$ & $(Z_6\times Z_2)\rtimes Z_2$  & $D_4$ ~~$(\tau\in \Sigma_2$)& $+$\\ \hline
6. &$\begin{pmatrix} \omega & 0 \\ 0 & i \end{pmatrix}$ & $Z_{12}\times Z_2$ & $Z_{12}\times Z_2$ ~~$(\tau\in \Sigma_1)$ & $+$\\ \hline\hline
\end{tabular} }
\caption{\label{FixedPoints} Inequivalent fixed points of $Sp(4,\mathbb{Z})$ in the Siegel upper half plane $\mathcal{H}_2$. We set $\zeta= e^{2\pi i /5},~\eta=\frac{1}{3}(1+i2\sqrt{2}),~\omega=-1/2+i\sqrt{3}/2$.
All other fixed points are related to one of these through a $Sp(4,\mathbbm{Z})$ transformation.
The group with \texttt{GAP} id [72,30] is isomorphic to
$(Z_6\times Z_6)\rtimes Z_2$. }
\end{table}

In modular invariant theories depending on a single modulus $\tau$ where lepton doublets are assigned to irreducible triplets of the finite modular group, it has been shown that in the vicinity of the fixed points
the predictions are universal: they are independent of the finite modular group acting on the lepton multiplets,
the weights of the matter multiplets and even the form of the kinetic terms.
We are led to study the behavior of Siegel modular invariant theories when the matrix $\tau$, conveniently restricted to $\Sigma_{2}$, falls in the vicinity of one of fixed points 2, 3, 4 and 5.
%%%%%%%%%%%%%%%%%%%%%%%%%%%%%%%%%%%%%%%%%%%%%%%%%%%%%%%%%
\section{Modular invariant models of lepton masses}
%%%%%%%%%%%%%%%%%%%%%%%%%%%%%%%%%%%%%%%%%%%%%%%%%%%%%%%%%
In the rest of this paper, we analyze CP-invariant and ${\cal N}=1$ locally supersymmetric models of lepton masses and mixing angles. Neglecting gauge interactions that are not of interest in our analysis, they are described by an action ${\mathscr S}$ fully specified by a gauge-invariant real function ${\cal G}$
\be
\label{eq:G-SUGRA}{\cal G}=\frac{K}{M_{Pl}^2}+\log \left\vert\frac{w}{M_{Pl}^3}\right\vert^2,
\ee
where restriction to the scalar components of the supermultiplets is understood, and $M_{Pl}$ is the Planck mass. Here $K$ is the K\"ahler potential, a real gauge-invariant function of the chiral multiplets and their conjugate of dimensionality (mass)$^2$, and $w$ is the superpotential, an analytic gauge-invariant function of the chiral multiplets of dimensionality (mass)$^3$. We do not discuss supersymmetry breaking effects, which in theories with a single modulus are negligible in a large portion of the parameter space~\cite{Criado:2018thu}. The chiral superfield content includes the matrix $\tau$ of eq.~\eqref{modulus}, restricted to the region $\Sigma_2$, and a set of matter fields $\varphi^{(I)}$. Siegel modular invariance requires ${\mathscr S}$ to remain unchanged under the transformations of eq.~\eqref{eq:tmg-1}, where $\gamma$ is restricted to the normalizer $N(H)$ of $\Sigma_{2}$. The transformation law of $\varphi^{(I)}$ is specified by a unitary representation $\rho_I(\gamma)$ of the finite modular group $N_n(H)$ and by the integer weight $k_I$ (irreducible components of $\varphi^{(I)}$ admitting independent weights). Invariance under CP requires ${\mathscr S}$ to remain unmodified under the transformations of eq.~\eqref{CP}.

Notice that for the action ${\mathscr S}$ to be invariant under the transformations of eq.~\eqref{eq:tmg-1}, the K\"ahler potential $K$ and the
superpotential $w$ do not need to be separately invariant. If modular transformations of eq.~\eqref{eq:tmg-1}, restricted to the scalar components $z$ of the supermultiplets $(\tau,\varphi^{(I)})$, induce a K\"ahler transformation:
\begin{align}
\label{KK}
K\xrightarrow{\gamma}&~K+M_{Pl}^2 F_\gamma(z)+M_{Pl}^2 \bar F_\gamma(\bar z)\,, \nn\\
w\xrightarrow{\gamma}&~e^{\dd -F_\gamma(z)} w\,,
\end{align}
${\cal G}$ does not change and the theory is invariant, provided the fermionic partners $\psi$ of the scalar multiplets
$z$ undergo an extra chiral rotation of the type~\footnote{Gauginos are required to transform with an opposite phase.}:
\be
\label{chiral}
\psi\xrightarrow{\gamma}e^{\frac{F_\gamma(z)-\bar F_\gamma(z)}{4}}\psi\,.
\ee
If $F_\gamma(z)$ has a nontrivial field dependence, eq.~\eqref{chiral} represents a local chiral rotation on the fermion components and anomaly
cancellation is required to guarantee invariance. Thus, up to potential anomalies requiring a cancellation mechanism, a K\"ahler transformation is always a symmetry of the theory. As a particular case, we can consider $F=-2 i\alpha_\gamma$, with $\alpha_\gamma$ a field-independent real constant
and eq.~\eqref{KK} becomes
\begin{align}
\label{KKa}
K\xrightarrow{\gamma}~K\,,~~~~~~w\xrightarrow{\gamma}~e^{\dd 2 i\alpha_\gamma} w\,.
\end{align}
For example, consider the choice:
\be
K=-3M_{Pl}^2 \log(-i\tau+i\bar\tau)\,,~~~~~~w=c\frac{M_{Pl}^3}{\eta(\tau)^6}\,,
\ee
where $\eta(\tau)$ is the Dedekind eta function and $c$ a constant.
When performing the transformation $T:\tau\to\tau+1$, we reproduce eq.~\eqref{KKa} with $\alpha_T=-\pi/4$,
thanks to the property
\be
\eta(\tau+1)=e^{\dd i\frac{\pi}{12}}~ \eta(\tau).
\ee
This theory is invariant under $T$~\footnote{Indeed, it is invariant under the full $SL(2,\mathbbm{Z})$.}, despite the presence of nontrivial phases in the transformation of the superpotential. We conclude that in the local case, the covariance of $w$ is only required up to a phase $2 \alpha_\gamma$. Since this phase depends on the group element $\gamma$, consistency with the group law requires~\footnote{We can replace this condition with a less restrictive one, where $v(\gamma)=e^{\dd 2 i\alpha_\gamma}$ is a multiplier system obeying:
\be
v(\gamma_1)v(\gamma_2)j(\gamma_1,\gamma_2\tau)^{-k_w}j(\gamma_2,\tau)^{-k_w}=v(\gamma_1\gamma_2)
j(\gamma_1\gamma_2,\tau)^{-k_w}\,,~~~~~~~~~j(\gamma,\tau)=\det(C\tau+D),
\ee
where $k_w$ is the weight of $w$. When $k$ is an integer, this condition becomes equivalent to the one in eq.~\eqref{cc}.}
\be
\label{cc}
\alpha_{\gamma_1}+\alpha_{\gamma_2}=\alpha_{\gamma_1\gamma_2}\,.
\ee
Here we will make use of the possibility in eq.~\eqref{KKa}, which is rarely exploited in the bottom-up approach. Working in the context of local supersymmetry, we consider the general case where $K$ and $w$ satisfy
\begin{align}
\label{wsieg}
K\xrightarrow{\gamma}&~K+k_w M^2_{Pl}\log\det(C\tau+D)+M^2_{Pl}k_w\log\det(C\bar\tau+D)\,,\nn\\
w\xrightarrow{\gamma}&~ \det(C\tau+D)^{-k_w} \bm{r}_s(\gamma) w,
\end{align}
where $\bm{r}_s$ is a singlet representation of $N_n(H)$, which can differ from the trivial one by a phase factor.  A suitable mechanism of anomaly cancellation, not affecting the analysis of the light fermion masses, is understood in the present analysis.

As a side remark, we stress that, in general, requiring $w$ to transform as a nontrivial singlet $\bm{r}_s$ is not possible in rigid supersymmetry, where
the theory depends separately on $K$ and $w$. Indeed, if $w$ acquires a phase, the invariance of the theory can be guaranteed by an $R$-symmetry.
The transformation in eq.~\eqref{KKa} should read:
\be
\label{wphase}
w(\theta)\xrightarrow{\gamma}  e^{\dd 2 i\alpha_\gamma} w(e^{\dd - i\alpha_\gamma}\theta),
\ee
which can be absorbed by a redefinition of the Grassmann measure
\begin{align}
\theta\xrightarrow{\gamma} &~e^{\dd +i \alpha_\gamma}~\theta\,, \nn\\
d^2\theta\xrightarrow{\gamma}&~ e^{\dd -2 i \alpha_\gamma}~d^2\theta\,.
\end{align}
When $w$ is a polynomial, the required $R$-invariance can be achieved if the $R$-charges of
the chiral multiplets add up to $+2$ in each term of $w$.
In a modular invariant theory, $w$ is not necessarily polynomial since one of these fields is typically a modular form $Y(\tau)$.
Due to the non-homogeneous dependence of $Y(\tau)$ on $\tau$, if we assign a nonvanishing $R$-charge to $\tau$,
$Y(\tau)$ will not possess a definite $R$-charge, in general. Thus, the only possible $R$-charge assignment of both
$\tau$ and $Y(\tau)$ is $R=0$. But this amounts to saying that the scalar component of $\tau$ does not transform
under the considered symmetry. This is what happens when dealing with a traditional flavour symmetry.
Specific combinations of modular transformations behave as traditional flavour symmetries, such as $S^2$
in $SL(2,\mathbbm{Z})$-invariant theories or transformations belonging to the center of the
metaplectic modular group, arising in the context of fractional weight modular forms~\cite{Liu:2020msy,Yao:2020zml,Almumin:2021fbk,Ishiguro:2023jqb}. In the following, we will not
consider these special cases, and we focus on nontrivial transformations of the moduli $\tau$.

We will exploit the above framework to describe masses, mixing angles, and phases in the lepton sector.
The matter multiplets $\varphi^{(I)}$ include two Higgs doublets $H_{u,d}$, electroweak doublets $L$, singlets $E^c$ and, when neutrino masses arise from the seesaw mechanism, heavy singlets $N^c$.
Assuming Majorana neutrino masses, the low-energy superpotential $w$ of the lepton sector reads:
\begin{align}
\label{starting2}
w=-\dd\frac{1}{2\Lambda_L} (H_u L)^T {\cal Y}(\tau) (H_u L)-E^{cT} {\cal Y}_e(\tau) (H_d L),
\end{align}
where the first term can arise from the seesaw mechanism when heavy singlets $N^c$ are integrated out.
To minimize the number of free parameters, $L$ will be assigned to an irreducible triplet of the finite modular group $N_n(H)$.
In the class of theories described by the action ${\mathscr S}$, the lepton mass matrices are obtained by combining the holomorphic contribution arising from the superpotential $w$ with the non-holomorphic data coming from the K\"ahler potential.
In general, there is a large freedom related to both the allowed transformation properties of the matter fields (level $n$, weight $k_I$ and representation $\rho_I$)
and the inefficiency of modular invariance in constraining the K\"ahler potential.
Moreover, the holomorphic contribution is expressed in terms of Siegel modular forms, whose knowledge
for generic levels and weights $(n,k)$ is limited. These obstacles can be largely overcome
if $\tau$ falls in the vicinity of a fixed point $\tau_0$. In this case,
even giving up the full power of modular invariance, a considerable amount of information about
lepton masses and mixing angles survives from the approximate invariance under CP and the stability group $G_0$.
%%%%%%%%%%%%%%%%%%%%%%%%%%%%%%%%%%%%%%%%%%%%%%%%%%%%%%%%%
\subsection{A field redefinition}
\label{afr}
%%%%%%%%%%%%%%%%%%%%%%%%%%%%%%%%%%%%%%%%%%%%%%%%%%%%%%%%%
\noindent
At the fixed points $\tau_0$, both CP and the stability group
$G_0$
are preserved. To establish CP conservation, it might be convenient to use a nonstandard definition of CP transformation $g\mathcal{CP}\equiv\gamma^{-1}\circ\mathcal{CP}$ on $\tau$ and the matter fields:
\begin{align}
\label{nsCP}
\tau\xrightarrow{g\mathcal{CP}}&~\tau_{CP}=\gamma^{-1}(-\bar\tau)\,,~~~~~~~~\gamma=\begin{pmatrix}
A_0 & B_0\\
C_0 & D_0
\end{pmatrix}\,,\nn\\
\varphi^{(I)} \xrightarrow{g\mathcal{CP}}&~\det(C^t_0\bar{\tau}+A_0^t)^{-k_I}\rho_I(\gamma^{-1}) X_I\;\bar\varphi^{(I)}~\,,
\end{align}
where $\gamma$ is an element of $Sp(4,\mathbbm{Z})$ fulfilling $\gamma\tau_0=-\bar\tau_0$, so that the fixed point $\tau_0$ is invariant under $g\mathcal{CP}$.
In general both CP and the stability group $G_0$ are nonlinearly realized on $(\tau,\varphi^{(I)})$ and it is preferable to move to a field basis $(u,\Phi^{(I)})$ where
CP and $G_0$ act linearly. We will choose a basis where $u(\tau_0)=0$, so that the group $G_0$ and CP are
unbroken at the origin of the field space, $u=0$. This allows us to adopt $u$ as an order parameter for the breaking of
$G_0$ and CP. The new basis can be defined by the transformation:
\begin{align}
\label{newb}
u=&~e^{\dd -i\alpha}(\tau-\tau_0)(\tau-\bar\tau_0)^{-1}\,,~~~~~~~~~\tau=(1-e^{\dd i\alpha} u)^{-1}(\tau_0-e^{\dd i\alpha}u\bar{\tau}_0)\,, \nn\\
\Phi^{(I)}=&~[\det(1-e^{\dd i\alpha}u)]^{k_I}\varphi^{(I)}\,,
\end{align}
where the phase $\alpha$ can be conveniently adjusted to simplify the action of CP.
Denoting by
\be
h_i=\left(
\begin{array}{cc}
A_i &B_i\\
C_i &D_i
\end{array}
\right)
\ee
the generators of the stability group $G_0$, $h_i\tau_0=\tau_0$,we find~\footnote{We used the identities
$
\det(C_i\tau+D_i)=\det(C_i\tau_0+D_i)\frac{\det(1-e^{ i\alpha}h_iu)}{\det(1-e^{i\alpha}u)}$ and $A_i-\tau_0 C_i=(\tau_0 C_i^t+D_i^t)^{-1}$.}
\begin{align}
u\xrightarrow{h_i}&~h_iu=(A_i-\tau_0 C_i)~u~(A_i-\bar\tau_0 C_i)^{-1}\,,\nn\\
\Phi^{(I)}\xrightarrow{h_i}&~\Omega_I(h_i)\Phi^{(I)}\,,~~~~\Omega_I(h_i)=\det(C_i\tau_0+D_i)^{-k_I}\rho_I(h_i)\,.
\end{align}
It is always possible to parametrize $u$ in such a way that this linear transformation is unitary.
Moreover, if $\rho_I(h_i)$ is unitary, so is $\Omega_I(h_i)$, since $[\det(C_i\tau_0+D_i)]^q=1$, where $q$ is the order of $h_i$.
The explicit parametrization of $u$ and $\Phi^{(I)}$ will be specified in the next Sections. Moreover, under the general CP symmetry $g\mathcal{CP}$, the new fields $u$ and $\Phi^{(I)}$ transform as follows
\begin{eqnarray}\label{eq:gCPtransf}
\nonumber u&\xrightarrow{g\mathcal{CP}}&e^{\dd-2i\alpha} \left(\bar{\tau}_0C_0+A_0\right)^{-1}\bar{u} \left(\tau_0C_0+A_0\right)\,,\\
\Phi^{(I)}&\xrightarrow{g\mathcal{CP}}&\det(\tau_0 C_0+A_0)^{k_I}X'_I\bar\Phi^{(I)},~~~X'_I=\rho_I(\gamma^{-1})X_{I}\,.
\end{eqnarray}
The phase $\det(\tau_0 C_0+A_0)^{k_I}$ can be absorbed by field redefinition further, thus the CP transformation matrix is essentially $X'_I$. The explicit form of $X'_I$ can be determined by solving a set of consistency conditions, as shown in the following Sections.
In the basis $(u,\Phi^{(I)})$, CP and $G_0$ are linearly realized in the field space, and are broken by the VEV of the flavon $u(\tau)$, which remains small if $\tau$ is near $\tau_0$.
Assuming that the lepton doublets $L$ transform as an irreducible triplet of $N_n(H)$, we can
build the most general set of lepton mass matrices with the correct transformation properties under $G_0$ and CP
from the decomposition of such a triplet under $G_0$.
As we will see, both the moduli $u$ and the irreducible triplets of $N_n(H)$ decompose into the direct sum of $G_0$-singlets. Thus, a suitable choice of the phase $\alpha$ allows us to cast the CP transformations of the relevant fields in the simple form:
\begin{align}
\label{eq:gCP-singlet-decom}\Phi^{(I)}\xrightarrow{g\mathcal{CP}}~\bar\Phi^{(I)},~~~~u_i\xrightarrow{g\mathcal{CP}}~\bar{u}_i\,.
\end{align}
Working in the vicinity of the fixed point, we only need a few terms of the expansion of such matrices in powers of the symmetry-breaking order parameter $u(\tau)$. In general, the kinetic terms originating from the most general K\"ahler potential allowed by CP and Siegel modular invariance are not canonical. After moving to the basis where such terms are canonical, it is not difficult to prove that, when the theory is invariant under $G_0$ and CP,
charged lepton and neutrino mass matrices, $m_e^2(u,\bar u)\equiv m_e(u,\bar u)^\dagger m_e(u,\bar u)$ and $m_\nu(u,\bar u)$, should transform as shown in table~\ref{constraints}.
\begin{table}[h!]
\centering
%\resizebox{1.0\textwidth}{!}{
\begin{tabular}{|lll|}
\hline\hline
& $G_0$& {\rm CP}\\
\hline
%\hline
$m_\nu(u,\bar u)$&$\bm{r}_s\Omega^*~m_\nu(u,\bar u)~\Omega^\dagger$&$m_\nu(u, \bar u)^*$\\
\hline
$m_\nu(u,\bar u)^{-1}$&$\bm{r}_s^{-1}\Omega~m_\nu(u,\bar u)^{-1}~\Omega^T$&$m_\nu(u, \bar u)^{-1*}$\\
\hline
$m_e(u,\bar u)$&$\bm{r}_s \Omega_c^*~m_e(u,\bar u)~\Omega^\dagger$&$m_e(u,\bar u)^*$\\
\hline
$m_{\bar e e}(u,\bar u)$&$\Omega~m_{\bar e e}(u,\bar u)~\Omega^\dagger$&$[m_{\bar e e}(u,\bar u)]^*$ \\
\hline\hline
\end{tabular}
%}
\caption{\label{constraints}Transformation properties of the lepton mass matrices under the group $G_0$ and CP. We have defined $m_{\bar e e}(u,\bar u)\equiv m_e(u,\bar u)^\dagger m_e(u,\bar u)$. In the column "$G_0$", we show the various mass matrices evaluated at $(h_i u, h_i\bar u)$, while in the column "CP" they are evaluated at $(\bar{u}_i, u_i)$. We allow for the possibility that the superpotential $w$ transforms as a generic singlet $\bm{r}_s$ of $N_n(H)$. }
\end{table}

The unitary matrices in table~\ref{constraints} read:
\begin{align}
\label{trans2}
\Omega\equiv \Omega_{H_u}\Omega_L\,,~~~~~~\Omega_c\equiv \Omega_{H_u}^*\Omega_{H_d}\Omega_{E^c}\,.
\end{align}
If the neutrino mass matrix arises from the seesaw mechanism, it may occur that $m_\nu(0,0)$ is singular~\footnote{That is the limit of $m_\nu(u,\bar u)$ when $u$ goes to zero does not exist or is infinite.}. In such a case it is convenient to enforce the transformations on the inverse $[m_\nu(u,\bar u)]^{-1}$, also reported in table~\ref{constraints}. Once $\Omega$ and $\Omega_c$ are specified, table~\ref{constraints} can be used to get the most general parametrization of $m_\nu(u,\bar u)$ and $m_{\bar ee}(u,\bar u)$
in the vicinity of $\tau_0$.
%%%%%%%%%%%%%%%%%%%%%%%%%%%%%%%%%%%%%%%%%%%%%%%%%%%%%%%%%%%%
\section{Siegel modular invariant models near the fixed points}
%%%%%%%%%%%%%%%%%%%%%%%%%%%%%%%%%%%%%%%%%%%%%%%%%%%%%%%%%%%%%
In this Section we analyze the fixed points 2, 3, 4 and 5, embedded in the region $\Sigma_{2}$, whose normalizer $N(H)$ is generated by the elements $G_i$ $(i=1,...,4)$ of eq.~\eqref{eq:NorTwo2}. We show that CP is conserved at each of these fixed points. We find out the stability group $G_0$ and its generators. Moving to the basis where $G_0$ acts linearly, we determine the transformation properties of moduli and fields under $G_0$ and CP. Finally, we find the decomposition of any irreducible triplet of the finite Siegel modular group $N_n(H)$ under the subgroup $G_0$.
%%%%%%%%%%%%%%%%%%%%%%%%%%%%%%%%%%%%%%%%%%%%%%%%%%%%%%%%%%%%
\subsection{Fixed point 2}
%%%%%%%%%%%%%%%%%%%%%%%%%%%%%%%%%%%%%%%%%%%%%%%%%%%%%%%%%%%%%
We start by analyzing the fixed point
\be
\label{t02}
\tau_0=
\begin{pmatrix} \eta & \frac{1}{2}(\eta -1) \\ \frac{1}{2}(\eta -1) & \eta \end{pmatrix}\,,~~~~~~~\eta=\frac{1}{3}(1+i2\sqrt{2})\,.
\ee
In $\tau_0$ CP is unbroken, since~\footnote{The matrix $\gamma$ is not unique: both $\gamma$ and $\gamma'=\gamma h$, with $h \tau_0=\tau_0$, satisfy eq.~\eqref{CP20}.}
\be
\label{CP20}
-\bar \tau_0=\gamma \tau_0\,,~~~~~~~
\gamma=\left(
\begin{array}{cccc}
1&0&0&0\\
0&1&0&0\\
-1&1&1&0\\
1&-1&0&1
\end{array}
\right)\,.
\ee
It is convenient to define the CP action on $\tau$ as in eq.~\eqref{nsCP}, with the matrix $\gamma$ of eq.~\eqref{CP20}, such that $\tau_{CP}$ belongs to the region conjugate to $\Sigma_2$ and $\tau_{0CP}=\tau_0$.
%%%%%%%%%%%%%%%%%%%%%%%%%%%%%%%%%%%%%%%%%%%%%%%%%%%%%%%%%%%%%
\subsubsection{Stability group}
%%%%%%%%%%%%%%%%%%%%%%%%%%%%%%%%%%%%%%%%%%%%%%%%%%%%%%%%%%%%%
The stability group $G_0$ of $\tau_0$, subgroup of the flavour group $N(H)$ defined in eq.~\eqref{eq:NorTwo2}, is isomorphic to $D_4$, generated by the elements $a$ and $b$, satisfying $a^4=b^2=(ab)^2=1$. In terms of the generators $G_i$ $(i=1,2,3,4)$ of $N(H)$, $a$ and $b$ read:
\be
\label{eq:gen-2nd-fixed}
a=G_4G_2G_3G_2\,,~~~~~~~~~b=(G_3G_2)^3\,.
\ee
The irreducible representations of $D_4$ are four singlets $\bm{1}_+$, $\bm{1}_-$, $\bm{1}'_+$, $\bm{1}'_-$ and one doublet $\bm{2}$.
Further details about $D_4$ are given in Appendix~\ref{app:groupD4}.
%%%%%%%%%%%%%%%%%%%%%%%%%%%%%%%%%%%%%%%%%%%%%%%%%%%%%%%%%%%%%
\subsubsection{New basis}
%%%%%%%%%%%%%%%%%%%%%%%%%%%%%%%%%%%%%%%%%%%%%%%%%%%%%%%%%%%%%
To realize a linear and unitary action of the stability group $D_4$ on $\tau$ and an antiunitary action of CP, we perform the field redefinition of eq.~\eqref{newb}, choosing
\be
e^{2 i\alpha}=-\frac13-i\frac{2\sqrt{2}}{3}\,.
\ee
Moreover, we use the parametrization:
\be
u=
\left(
\begin{array}{cc}
\frac{1}{\sqrt{3}}u_1+i\sqrt{\frac23}u_3&i\sqrt{\frac23}u_1+\frac{1}{\sqrt{3}}u_3\\
i\sqrt{\frac23}u_1+\frac{1}{\sqrt{3}}u_3&\frac{1}{\sqrt{3}}u_1+i\sqrt{\frac23}u_3
\end{array}
\right)\,.
\ee
The phase $\alpha$ reveals useful in simplifying the action of CP on $u_{1,3}$.
Under the stability group $D_4$, the new fields $u$ split into the sum of an invariant singlet, $u_1$, and a component $u_3$ transforming as $\bm{1}'_-$, see Appendix~\ref{app:groupD4}:
\begin{align}
&u_1\xrightarrow{a}+u_1,\qquad u_1\xrightarrow{b}+u_1,\qquad u_1\sim \bm{1}_+\nn\\
&u_3\xrightarrow{a}-u_3,\qquad u_3\xrightarrow{b}+u_3,\qquad u_3\sim \bm{1}'_-\,.
\end{align}
Under CP, we get
\be
\label{eq:u-gCP-FP2}u_1\xrightarrow{g\mathcal{CP}} \bar u_1\,,~~~~~~~~~u_3\xrightarrow{g\mathcal{CP}} \bar u_3\,.
\ee
The action of $D_4$ on the new field $\Phi^{(I)}$ of eq.~\eqref{newb} reads
\begin{align}
\Phi^{(I)}\xrightarrow{\gamma}\Omega_I(\gamma)\Phi^{(I)}\,,~~~~~~~
\Omega_I(\gamma)=j(\gamma,\tau_0)^{-k_I}\rho_I(\gamma)\,.
\end{align}
\noindent
where $j(\gamma,\tau)=\det(C\tau+D)$.
In particular, we have:
\be
j(a,\tau_0)^{-1}=+1\,,~~~~~~j(b,\tau_0)^{-1}=-1,
\ee
which reproduces the representation $\bm{1}_-$ of the stability group. We see that $\Omega_I(\gamma)=j(\gamma,\tau_0)^{-k_I}\rho_I(\gamma)$ is also a representation of the stability group, the direct product of $(\bm{1}_-)^{k_I}=(\bm{1}_-,\bm{1}_+)$ and $\rho_I(\gamma)$.
In a bottom-up approach, by varying all possible representations of the matter multiplets, we can absorb the factor $(\bm{1}_-)^{k_I}$ into the choice of $\rho_I(\gamma)$. From Eq.~\eqref{eq:gCP-cons-cond} and the expression of generators in eq.~\eqref{eq:gen-2nd-fixed}, we know the CP transformation $X'_I$ satisfies the following consistency conditions on the stability group $D_4$:
\begin{equation}
\label{eq:gCP-cons-FP2}X'_I\,\rho^{*}_I(a)X'^{-1}_I=\rho_I(a),~~~~X'_I\,\rho^{*}_I(b)X'^{-1}_I=\rho_I(b^{-1})\,,
\end{equation}
which lead to
\begin{equation}
X'_I=\mathbbm{1}
\end{equation}
in the basis of Appendix~\ref{app:groupD4}, no matter whether $\Phi^{(I)}$ transforms as singlets or doublet of $D_4$. Hence the CP transformation simply maps $\Phi^{(I)}$ to its conjugate.

%%%%%%%%%%%%%%%%%%%%%%%%%%%%%%%%%%%%%%%%%%%%%%%%%%%%%%%%%%%%%
\subsubsection{Triplet decomposition}
%%%%%%%%%%%%%%%%%%%%%%%%%%%%%%%%%%%%%%%%%%%%%%%%%%%%%%%%%%%%%
From the 168 triplet representation of the finite Siegel modular groups $N_n(H)$, the three-dimensional representation matrices of the $D_4$ generators $a$ and $b$ can be obtained via the relations in Eq.~\eqref{eq:gen-2nd-fixed}.
This allows the decomposition of the three-dimensional irreducible representations of $N_n(H)$ under the stability group $D_4$. We find that each irreducible triplet $ \rho_L$ decomposes into one of the following sum of three singlets of $D_4$
\begin{equation}
\label{eq:4decomposition1}
\rho_L \sim \begin{cases}
\mathbf{1}_{-} \oplus \mathbf{1}^\prime_{+} \oplus \mathbf{1}^\prime_{+} \\
\mathbf{1}_{+} \oplus \mathbf{1}^\prime_{-} \oplus \mathbf{1}^\prime_{-}  \\
\mathbf{1}^\prime_{+} \oplus \mathbf{1}_{-} \oplus \mathbf{1}_{-} \\
\mathbf{1}^\prime_{-} \oplus \mathbf{1}_{+} \oplus \mathbf{1}_{+} \\
\end{cases}\,.
\end{equation}
We end up with
\be
\Omega_L(\gamma)=j(\gamma,\tau_0)^{-k_I}\rho_L(\gamma)\,,
\ee
where $j(\gamma,\tau_0)^{-1}\sim\bm{1}_-$. Finally, we get $\Omega\equiv \Omega_{H_u}\Omega_L$ by multiplying $\Omega_L$ and the generic singlet associated with $\Omega_{H_u}$.

%%%%%%%%%%%%%%%%%%%%%%%%%%%%%%%%%%%%%%%%%%%%%%%%%%%%%%%%%%%%
\subsection{Fixed point 3}
%%%%%%%%%%%%%%%%%%%%%%%%%%%%%%%%%%%%%%%%%%%%%%%%%%%%%%%%%%%%%
The fixed point
\be
\label{t0}
\tau_0=
\left(
\begin{array}{cc}
i&0\\0&i
\end{array}
\right)
\ee
belongs to both $\Sigma_1$ and $\Sigma_2$. It preserves CP using the standard definition of eq.~\eqref{nsCP}, where $\gamma=\mathbbm{1}$. We assume that in the theory under examination, the matrix $\tau$ is restricted to $\Sigma_{2}$ and the flavor group is the corresponding normalizer $N(H)$ defined in eq.~\eqref{eq:NorTwo2}.
%%%%%%%%%%%%%%%%%%%%%%%%%%%%%%%%%%%%%%%%%%%%%%%%%%%%%%%%%%%%%
\subsubsection{Stability group}
%%%%%%%%%%%%%%%%%%%%%%%%%%%%%%%%%%%%%%%%%%%%%%%%%%%%%%%%%%%%%
The stability group $G_0$ of $\tau_0$, subgroup of the flavour group $N(H)$ defined in eq.~\eqref{eq:NorTwo2}, is isomorphic to the Pauli group, namely the central product of $D_4$ and $Z_4$: $D_4\circ Z_4$. $G_0$ is generated by the elements $a$ and $b$ and $c$, satisfying $a^4=c^4=b^2=(ab)^2=1$, $a^2=c^2$, $ac=ca$, $bc=cb$. The elements $(a,b)$ generate $D_4$ and $c$ generates $Z_4$. The elements of $D_4$ commute with those of $Z_4$. In terms of the generators $G_i$ $(i=1,2,3,4)$ of $N(H)$, $a$, $b$ and $c$ read:
\be
a=G_3G_4\,,~~~~~~b=(G_3G_2)^3\,,~~~~~~c=G_3\,.
\ee
The irreducible representations of $D_4\circ Z_4$ are eight singlets and two doublets. Further details about $D_4\circ Z_4$ are given in Appendix~\ref{app:groupPauli}.
%%%%%%%%%%%%%%%%%%%%%%%%%%%%%%%%%%%%%%%%%%%%%%%%%%%%%%%%%%%%%
\subsubsection{New basis}
%%%%%%%%%%%%%%%%%%%%%%%%%%%%%%%%%%%%%%%%%%%%%%%%%%%%%%%%%%%%%
We move to the new basis of eq.~\eqref{newb}, choosing $\alpha=0$.
Using the parametrization:
\be
u=
\left(
\begin{array}{cc}
u_1~&~u_3\\
u_3~&~u_1
\end{array}
\right)\,,
\ee
we find that, under the stability group $D_4\circ Z_4$, the new multiplet $u$ splits into the sum of two singlets: $\bm{1}_{-+-}\oplus\bm{1}_{++-}$, see Appendix~\ref{app:groupPauli}:
\begin{align}
&u_1\xrightarrow{a}-u_1\,,\qquad u_1\xrightarrow{b}+u_1\,,\qquad u_1\xrightarrow{c}-u_1\,,\qquad u_1\sim \bm{1}_{-+-}\nn\\
&u_3\xrightarrow{a}+u_3\,,\qquad u_3\xrightarrow{b}+u_3\,,\qquad u_3\xrightarrow{c}-u_3\,,\qquad u_3\sim \bm{1}_{++-}~\,.
\end{align}
Under CP, with $\gamma=\mathbbm{1}$ in eq.~\eqref{nsCP}, we get
\be
u_1\xrightarrow{\mathcal{CP}} \bar u_1\,,~~~~~~~~~~u_3\xrightarrow{\mathcal{CP}} \bar u_3\,.
\ee
The action of $D_4\circ Z_4$ on the new matter fields $\Phi^{(I)}$ of eq.~\eqref{newb} is
\begin{align}
\Phi^{(I)}\xrightarrow{\gamma}\Omega_I(\gamma)\Phi^{(I)}\,,~~~~~~
\Omega_I(\gamma)=j(\gamma,\tau_0)^{-k_I}\rho_I(\gamma)\,.
\end{align}
where $j(\gamma,\tau)=\det(C\tau+D)$. In particular, we have:
\be
j(a,\tau_0)^{-1}=+1\,,~~~~~~j(b,\tau_0)^{-1}=-1\,,~~~~~~j(c,\tau_0)^{-1}=-1\,,
\ee
which reproduces the representation $\bm{1}_{+--}$ of the stability group. We see that $\Omega_I(\gamma)=j(\gamma,\tau_0)^{-k_I}\rho_I(\gamma)$ is also a representation of the stability group, the direct product of  $(\bm{1}_{+--})^{k_I}$ and $\rho_I(\gamma)$.
In a bottom-up approach, by varying all possible representations of the matter multiplets, we can absorb the factor $(\bm{1}_{+--})^{k_I}$ into the choice of $\rho_I(\gamma)$. The consistency conditions of CP transformation $X'_I$ on the stability group $D_4\circ Z_4$ is given by
\begin{equation}
\label{eq:gCP-cons-FP3}X'_I\,\rho^{*}_I(a)X'^{-1}_I=\rho_I(a^{-1}),~~~~X'_I\,\rho^{*}_I(b)X'^{-1}_I=\rho_I(b^{-1}),~~~~X'_I\,\rho^{*}_I(c)X'^{-1}_I=\rho_I(c^{-1})\,.
\end{equation}
In the basis of Appendix~\ref{app:groupPauli}, we have $X'_I=\mathbbm{1}$ regardless of the transformation of $\Phi^{(I)}$ under $D_4\circ Z_4$. As a result, the CP transformation of the matter field is $\Phi^{(I)}\xrightarrow{\mathcal{CP}}\bar\Phi^{(I)}$.

%%%%%%%%%%%%%%%%%%%%%%%%%%%%%%%%%%%%%%%%%%%%%%%%%%%%%%%%%%%%%
\subsubsection{Triplet decomposition}
%%%%%%%%%%%%%%%%%%%%%%%%%%%%%%%%%%%%%%%%%%%%%%%%%%%%%%%%%%%%%
We find that each irreducible triplet of $N_n(H)$
decomposes under $D_4\circ Z_4$ into a direct sum of three singlets, wherein two of them are identical. There are only eight distinct cases as follows,
\begin{align}
\label{eq:8decomposition-case3}
\rho_L \sim \begin{cases}
\bm{1}_{+++} \oplus \bm{1}_{-+-} \oplus \bm{1}_{-+-} \\
\bm{1}_{-+-} \oplus \bm{1}_{+++} \oplus \bm{1}_{+++} \\
\bm{1}_{+--} \oplus \bm{1}_{--+} \oplus \bm{1}_{--+} \\
\bm{1}_{--+} \oplus \bm{1}_{+--} \oplus \bm{1}_{+--} \\
\bm{1}_{+-+} \oplus \bm{1}_{---} \oplus \bm{1}_{---}\\
\bm{1}_{---} \oplus \bm{1}_{+-+} \oplus \bm{1}_{+-+}\\
\bm{1}_{-++} \oplus \bm{1}_{++-} \oplus \bm{1}_{++-} \\
\bm{1}_{++-} \oplus \bm{1}_{-++} \oplus \bm{1}_{-++}
\end{cases}\,.
\end{align}
We end up with
\be
\Omega_L(\gamma)=j(\gamma,\tau_0)^{-k_I}\rho_L(\gamma)\,,
\ee
where $j(\gamma,\tau_0)^{-1}\sim\bm{1}_{+--}$. Finally, we get $\Omega\equiv \Omega_{H_u}\Omega_L$
by multiplying $\Omega_L$ and the generic singlet associated with $\Omega_{H_u}$.
%%%%%%%%%%%%%%%%%%%%%%%%%%%%%%%%%%%%%%%%%%%%%%%%%%%%%%%%%%%%
\subsection{Fixed point 4}
%%%%%%%%%%%%%%%%%%%%%%%%%%%%%%%%%%%%%%%%%%%%%%%%%%%%%%%%%%%%%
The fixed point
\be
\label{t0}
\tau_0=
\left(
\begin{array}{cc}
\omega&0\\0&\omega
\end{array}
\right)
\ee
belongs to both $\Sigma_1$ and $\Sigma_2$. It preserves CP using the definition of eq.~\eqref{nsCP}, with
\be
\label{gamma4}
\gamma=\left(
\begin{array}{cccc}
1&0&1&0\\
0&1&0&1\\
0&0&1&0\\
0&0&0&1
\end{array}
\right).
\ee
We assume that in the theory under examination the matrix $\tau$ is restricted to $\Sigma_{2}$ and the flavor group is the corresponding normalizer $N(H)$ defined in eq.~\eqref{eq:NorTwo2}.
%%%%%%%%%%%%%%%%%%%%%%%%%%%%%%%%%%%%%%%%%%%%%%%%%%%%%%%%%%%%%
\subsubsection{Stability group}
%%%%%%%%%%%%%%%%%%%%%%%%%%%%%%%%%%%%%%%%%%%%%%%%%%%%%%%%%%%%%
The stability group $G_0$ of $\tau_0$, subgroup of the flavour group $N(H)$ defined in eq.~\eqref{eq:NorTwo2}, is isomorphic to  $D_4\times Z_3$ generated by the elements $a$ and $b$ and $c$, satisfying $a^4=b^2=(ab)^2=c^3=1$, $ca=ac$, $cb=bc$. The elements $(a,b)$ generate $D_4$ and $c$ generates $Z_3$. In terms of the generators $G_{i}$, $(i=1,2,3,4)$ of $N(H)$, eq.~\eqref{eq:NorTwo2}, $a$, $b$ and $c$ read:
\be
a=(G_2G_3)^3G_4\,,~~~~~~~b=G_3^2G_4\,,~~~~~~~c=(G_3G_1)^2.
\ee
The irreducible representations of $D_4\times Z_3$ are 12 singlets and 3 doublets. Further details about $D_4\times Z_3$ are given in Appendix~\ref{app:groupD4Z3}.
%%%%%%%%%%%%%%%%%%%%%%%%%%%%%%%%%%%%%%%%%%%%%%%%%%%%%%%%%%%%%
\subsubsection{New basis}
%%%%%%%%%%%%%%%%%%%%%%%%%%%%%%%%%%%%%%%%%%%%%%%%%%%%%%%%%%%%%
We move to the new basis of eq.~\eqref{newb}, choosing $\alpha=0$.
Using the parametrization:
\be
u=
\left(
\begin{array}{cc}
u_1&u_3\\
u_3&u_1
\end{array}
\right)\,,
\ee
we find that, under the stability group $G_0=D_4\times Z_3$, the new multiplet $u$ transforms as the sum of two singlets: $u_1\sim\bm{1}_{++1}$ and $u_3\sim\bm{1}_{--1}$, see Appendix~\ref{app:groupD4Z3}.
\begin{align}
&u_1\xrightarrow{a}+u_1\,,\qquad u_1\xrightarrow{b}+u_1\,,\qquad u_1\xrightarrow{c}\omega u_1\nn\\
&u_3\xrightarrow{a}-u_3\,,\qquad u_3\xrightarrow{b}-u_3\,,\qquad u_3\xrightarrow{c}\omega u_3\,,
\end{align}
where $\omega=-1/2+i\sqrt{3}/2$. Under CP, with $\gamma$ in eq.~\eqref{gamma4}, we get
\be
u_1\xrightarrow{g\mathcal{CP}}\bar{u}_1\,,~~~~~~~u_2\xrightarrow{g\mathcal{CP}} \bar{u}_2\,.
\ee
The action of $D_4\times Z_3$ on the new matter fields $\Phi^{(I)}$ of eq.~\eqref{newb} is
\begin{align}
\Phi^{(I)}\xrightarrow{\gamma}\Omega_I(\gamma)\Phi^{(I)}\,,~~~~~~~~~
\Omega_I(\gamma)=j(\gamma,\tau_0)^{-k_I}\rho_I(\gamma)\,.
\end{align}
\noindent
where $j(\gamma,\tau)=\det(C\tau+D)$. In particular, we have:
\be
j(a,\tau_0)^{-1}=1\,,~~~~~~j(b,\tau_0)^{-1}=-1\,,~~~~~~j(c,\tau_0)^{-1}=\omega\,,
\ee
which reproduces the representation $\bm{1}_{+-1}$ of the stability group. We see that $\Omega_I(\gamma)=j(\gamma,\tau_0)^{-k_I}\rho_I(\gamma)$ is also a representation of the stability group, the direct product of  $(\bm{1}_{+-1})^{k_I}$ and $\rho_I(\gamma)$.
In a bottom-up approach, by varying all possible representations of the matter multiplets, we can absorb the factor $(\bm{1}_{+-1})^{k_I}$ into the choice of $\rho_I(\gamma)$. The CP transformation $X'_I$ fulfills the following consistency conditions on the stability group $D_4\times Z_3$:
\begin{equation}
\label{eq:gCP-cons-FP4}X'_I\,\rho^{*}_I(a)X'^{-1}_I=\rho_I(a)\,,~~~~X'_I\,\rho^{*}_I(b)X'^{-1}_I=\rho_I(b^{-1})\,,~~~~X'_I\,\rho^{*}_I(c)X'^{-1}_I=\rho_I(c^{-1})\,.
\end{equation}
A basis for the $D_4\times Z_3$ generators exists, where the consistency conditions of CP transformation restricted to the stability group $D_4\times Z_3$ are solved by $X_I'=\mathbbm{1}$, see comment at the end of Appendix \ref{app:groupD4Z3}.

%%%%%%%%%%%%%%%%%%%%%%%%%%%%%%%%%%%%%%%%%%%%%%%%%%%%%%%%%%%%%
\subsubsection{Triplet decomposition}
%%%%%%%%%%%%%%%%%%%%%%%%%%%%%%%%%%%%%%%%%%%%%%%%%%%%%%%%%%%%%
We find that each irreducible triplet of $N_n(H)$
decomposes under $D_4\times Z_3$ into a direct sum of three singlets which transform in the same way under the $D_4\times Z_3$ subgroup. There are only four distinct cases as follows,
\begin{align}
\label{eq:8decomposition-case3}
\rho_L \sim \begin{cases}
\bm{1}_{++0} \oplus \bm{1}_{++1} \oplus \bm{1}_{++2} \\
\bm{1}_{+-0} \oplus \bm{1}_{+-1} \oplus \bm{1}_{+-2} \\
\bm{1}_{-+0} \oplus \bm{1}_{-+1} \oplus \bm{1}_{-+2} \\
\bm{1}_{--0} \oplus \bm{1}_{--1} \oplus \bm{1}_{--2}
\end{cases}\,.
\end{align}
We end up with
\be
\Omega_L(\gamma)=j(\gamma,\tau_0)^{-k_I}\rho_L(\gamma)\,,
\ee
where $j(\gamma,\tau_0)^{-1}\sim\bm{1}_{+-1}$. Finally, we get $\Omega\equiv \Omega_{H_u}\Omega_L$ by multiplying $\Omega_L$ and the generic singlet associated with $\Omega_{H_u}$. Since each triplet of $N_n(H)$ is decomposed into three singlets of $D_4\times Z_3$, the CP symmetry transforms the lepton fields into their conjugate.
%%%%%%%%%%%%%%%%%%%%%%%%%%%%%%%%%%%%%%%%%%%%%%%%%%%%%%%%%%%%
\subsection{Fixed point 5}
%%%%%%%%%%%%%%%%%%%%%%%%%%%%%%%%%%%%%%%%%%%%%%%%%%%%%%%%%%%%%
The last case to be examined is the fixed point
\be
\label{t0}
\tau_0=\frac{i}{\sqrt{3}}
\left(
\begin{array}{cc}
2&1\\1&2
\end{array}
\right).
\ee
It preserves CP using the standard definition of eq.~\eqref{nsCP}, where $\gamma=\mathbbm{1}$.
%%%%%%%%%%%%%%%%%%%%%%%%%%%%%%%%%%%%%%%%%%%%%%%%%%%%%%%%%%%%%
\subsubsection{Stability group}
%%%%%%%%%%%%%%%%%%%%%%%%%%%%%%%%%%%%%%%%%%%%%%%%%%%%%%%%%%%%%
The stability group $G_0$ of $\tau_0$, subgroup of the flavour group $N(H)$ defined in eq.~\eqref{eq:NorTwo2}, is isomorphic to $D_4$, generated by the elements $a$ and $b$, satisfying $a^4=b^2=(ab)^2=1$. In terms of the generators $G_i$ $(i=1,2,3,4)$ of $N(H)$, $a$ and $b$ read:
\be
a=G_3G_4\,,~~~~~~~b=(G_3G_2)^3\,.
\ee
The irreducible representations of $D_4$ are four singlets $\bm{1}_+$, $\bm{1}_-$, $\bm{1}'_+$, $\bm{1}'_-$ and one doublet $\bm{2}$.
Further details about $D_4$ are given in Appendix~\ref{app:groupD4}.
%%%%%%%%%%%%%%%%%%%%%%%%%%%%%%%%%%%%%%%%%%%%%%%%%%%%%%%%%%%%%
\subsubsection{New basis}
%%%%%%%%%%%%%%%%%%%%%%%%%%%%%%%%%%%%%%%%%%%%%%%%%%%%%%%%%%%%%
We move to the new basis of eq.~\eqref{newb}, choosing $\alpha=0$.
We define:
\be
u=\left(
\begin{array}{cc}
u_1 & u_3\\
u_3 & u_1
\end{array}
\right)\,,
\ee
and we find that, under the stability group $D_4$, the new fields $u$ split into the sum of an invariant singlet, $u_3$, and a component $u_1$
transforming as $\bm{1}'_-$, see Appendix~\ref{app:groupD4}:
\begin{align}
&u_1\xrightarrow{a}-u_1\,,\qquad u_1\xrightarrow{b}+u_1\,,\qquad u_1\sim \bm{1}'_-\nn\\
&u_3\xrightarrow{a}+u_3\,,\qquad u_3\xrightarrow{b}+u_3\,,\qquad u_3\sim \bm{1}_+\,.
\end{align}
Under CP, we get
\be
u_1\xrightarrow{\mathcal{CP}} \bar u_1\,,~~~~~~~~u_3\xrightarrow{\mathcal{CP}} \bar u_3\,.
\ee
The action of $D_4$ on the new matter fields $\Phi^{(I)}$ of eq.~\eqref{newb} is
\begin{align}
\Phi^{(I)}\xrightarrow{\gamma}\Omega_I(\gamma)\Phi^{(I)}\,,~~~~~~~~~
\Omega_I(\gamma)=j(\gamma,\tau_0)^{-k_I}\rho_I(\gamma).
\end{align}
\noindent
where $j(\gamma,\tau)=\det(C\tau+D)$.
In particular, we have:
\be
j(a,\tau_0)^{-1}=+1\,,~~~~~~j(b,\tau_0)^{-1}=-1\,,
\ee
which reproduces the representation $\bm{1}_-$ of the stability group. We see that $\Omega_I(\gamma)=j(\gamma,\tau_0)^{-k_I}\rho_I(\gamma)$ is also a representation of the stability group, the direct product of $(\bm{1}_-)^{k_I}=(\bm{1}_-,\bm{1}_+)$ and $\rho_I(\gamma)$.
In a bottom-up approach, by varying all possible representations of the matter multiplets, we can absorb the factor $(\bm{1}_-)^{k_I}$ into the choice of $\rho_I(\gamma)$. The consistency conditions of CP transformation $X'_I$ on the stability group $D_4$ is given by
\begin{equation}
\label{eq:gCP-cons-FP5}X'_I\,\rho^{*}_I(a)X'^{-1}_I=\rho_I(a^{-1}),~~~~X'_I\,\rho^{*}_I(b)X'^{-1}_I=\rho_I(b^{-1})\,.
\end{equation}
A basis for the $D_4$ generators exists, where the consistency conditions of CP transformation restricted to the stability group $D_4$ are solved by $X_I'=\mathbbm{1}$, see comment at the end of Appendix \ref{app:groupD4}.
Notice that this basis does not coincide with the one allowing to choose $X_I'=\mathbbm{1}$ at the fixed point 2,
despite the stability group for the fixed points 2 and 5 is the same.
%%%%%%%%%%%%%%%%%%%%%%%%%%%%%%%%%%%%%%%%%%%%%%%%%%%%%%%%%%%%%
\subsubsection{Triplet decomposition}
%%%%%%%%%%%%%%%%%%%%%%%%%%%%%%%%%%%%%%%%%%%%%%%%%%%%%%%%%%%%%
We need the decomposition of the irreducible triplets $\rho_L$ of $N_n(H)$, to which lepton doublets $L$ are assigned, within the stability group $D_4$.
With the help of Appendix~\ref{app:groupD4}, by considering the representation matrices of the two generators, $a$ and $b$, we find that each irreducible triplet of $N_n(H)$ decomposes under $D_4$ into a direct sum of three singlets, wherein two of them are identical. There are only four distinct cases, illustrated in eq.~\eqref{eq:4decomposition}.
\begin{align}
\label{eq:4decomposition}
\rho_L \sim \begin{cases}
\mathbf{1}_{-} \oplus \mathbf{1}^\prime_{+} \oplus \mathbf{1}^\prime_{+} \\
\mathbf{1}_{+} \oplus \mathbf{1}^\prime_{-} \oplus \mathbf{1}^\prime_{-}  \\
\mathbf{1}^\prime_{+} \oplus \mathbf{1}_{-} \oplus \mathbf{1}_{-} \\
\mathbf{1}^\prime_{-} \oplus \mathbf{1}_{+} \oplus \mathbf{1}_{+} \\
\end{cases}\,,
\end{align}
We end up with
\be
\Omega_L(\gamma)=j(\gamma,\tau_0)^{-k_I}\rho_L(\gamma)\,,
\ee
where $j(\gamma,\tau_0)^{-1}\sim\bm{1}_-$. Finally, we get $\Omega\equiv \Omega_{H_u}\Omega_L$
by multiplying $\Omega_L$ and the generic singlet associated with $\Omega_{H_u}$.
%%%%%%%%%%%%%%%%%%%%%%%%%%%%%%%%%%%%%%%%%%%%%%%
\section{\label{sec:MassMatrix}Neutrino Mass Matrix}
%%%%%%%%%%%%%%%%%%%%%%%%%%%%%%%%%%%%%%%%%%%%%%%
In this Section, we discuss the patterns of the lepton mass matrices
of a generic theory whose moduli, belonging to the region $\Sigma_2$ parameterized by
\be
\tau=\begin{pmatrix} \tau_1 & \tau_3 \\ \tau_3 & \tau_1 \end{pmatrix}
\ee
are close to one of the fixed points examined previously. We assume the most general CP-invariant ${\cal N}=1$ local supersymmetric action and enforce Siegel modular invariance by asking that the superpotential $w$ transforms as in eq.~\eqref{wsieg} with $\bm{r}_s$ either trivial or non-trivial singlet of the finite modular group $N_n(H)$, and thus also a singlet of the stability group $G_0$. In such a theory the level $n$, the weights $k_I$ and the representations $\rho_I$ of the matter multiplets, and even the specific form of the K\"ahler potential are the most general ones consistent with the requirement of CP and modular invariance. In this completely general framework, we make a single assumption: the lepton doublets $L$ are assigned to an irreducible triplet $\rho_L$ of the finite modular group $N_n(H)$. In a bottom-up approach this assumption usually minimizes the number of free parameters of the theory.

Working with the local coordinates $(u,\Phi^{(I)})$ of eq.~\eqref{newb}, on which the stability group $G_0$ has a linear action, the lepton mass matrices $m_{\bar{e}e}$ and $m_\nu$ must transform as required by table~\ref{constraints}. Making use of the decomposition of each irreducible triplet of $N_n(H)$ and the transformation laws of $u$ under $G_0$, we can easily fulfill this requisite, which can be implemented order by order in the expansion in powers of $|u|$, a small quantity when $\tau$ is close to the fixed point. We provide the general expression of $m_{\bar{e}e}$ and $m_\nu$ at the first nontrivial order in $|u|$.

As a general result, we find that in each fixed point, the mass matrices $m_{\bar{e}e}$ and $m_\nu$ are sensitive only to $\bm{r}_s$ and not to the decomposition of $\rho_L$ under $G_0$. Moreover, the mass matrices $m_{\bar{e}e}$ are the same for all $\bm{r}_s$, while the neutrino mass matrices $m_\nu$(or $m_\nu^{-1}$) depend on $\bm{r}_s$. The examined fixed points fall into two classes. The fixed points 2, 3 and 5 give rise to similar patterns, which are different from those arising around the fixed point 4.
%%%%%%%%%%%%%%%%%%%%%%%%%%%%%%%%%%%%%%%%%%%%%%%
\subsection{Fixed points 2, 3, and 5}
%%%%%%%%%%%%%%%%%%%%%%%%%%%%%%%%%%%%%%%%%%%%%%%
From table~\ref{constraints} and the analysis of the previous Section, we find the following pattern for $m_{\bar e e}(u,\bar u)$:
\begin{align}
m_{\bar e e}(u,\bar u)=m_{0e}^2~
\left(
\begin{array}{ccc}
y^0_{11} &y_{12} ~x&y_{13} ~x\\
y_{12}^* ~x&y^0_{22}&y^0_{23} \\
y_{13}^* ~x&y^0_{23}&y^0_{33}\\
\end{array}
\right)+\ldots\,,
\end{align}
where $x=|u_3|$ for the fixed point 2 and $x=|u_1|$ for the fixed points 3 and 5. The overall real coefficient $m_{0e}^2$ has the dimension of (mass)$^2$. Dots stand for higher orders in the $x$ expansion. The coefficients $y^0_{ij}(y_{ij})$ are real (complex) numbers, independent of the moduli. They are not constrained in the present analysis, though they are expected to be of the same order. To discuss the lepton mixing matrix, we diagonalize $m_{\bar e e}(u,\bar u)$:
\begin{align}
U_e^\dagger m_{\bar e e}(u,\bar u) U_e={\tt diag} [m_{\bar e e}(u,\bar u)]\,.
\end{align}
Up to a permutation matrix $P$ related to the ordering of the charged lepton masses, $U_e$ has the pattern:
\begin{align}
\label{ue}
U_e=\left(
\begin{array}{ccc}
{\cal O}(1)&{\cal O}(x)&{\cal O}(x)\\
{\cal O}(x)&{\cal O}(1)&{\cal O}(1)\\
{\cal O}(x)&{\cal O}(1)&{\cal O}(1)
\end{array}
\right)\,.
\end{align}
Next, we examine the neutrino mass matrices. We find that they fall into three classes, depending on the singlet representation $\bm{r}_s$ under which $w$ transforms, see table~\ref{numasses}.
\begin{itemize}
\item[$\bullet$ {\bf A}] The following pattern is obtained when $\bm{r}_s$ embeds one of the $G_0$ singlets listed in table~\ref{numasses}, first line. The overall factor $m_{0\nu}$ is a real mass parameter.
\vskip 0.2 cm
\noindent
\begin{align}
\label{mA}
m_\nu(u,\bar u)=m_{0\nu}~
\left(
\begin{array}{ccc}
x^0_{11} &x_{12}~ x&x_{13} ~x\\
\cdot&x^0_{22}&x^0_{23} \\
\cdot&\cdot&x^0_{33}\\
\end{array}
\right)+\ldots\,.
\end{align}
The coefficients $x^0_{ij}(x_{ij})$ are real (complex) numbers, independent of the moduli $u_{1,3}$,
except for the choice $\bm{1}_{++-}$ at the fixed point 3, where both $x^0_{ij}$ and $x_{ij}$ are proportional to a linear combination of $u_3$ and $\bar u_3$. In any case, $x^0_{ij}(x_{ij})$ are expected to be of the same order, and we do not need to treat separately this special case.
\item[$\bullet$ {\bf B}] The following pattern is obtained when $\bm{r}_s$ embeds one of the $G_0$ singlets listed in table~\ref{numasses}, second line. The overall factor $m_{0\nu}$ is a real mass parameter.
\vskip 0.2 cm
\noindent
\begin{align}
\label{mB}
m_\nu(u,\bar u)=m_{0\nu}~
\left(
\begin{array}{ccc}
x_{11} ~x &x^0_{12} &x^0_{13} \\
\cdot&x_{22}~ x&x_{23}~ x \\
\cdot&\cdot&x_{33}~ x\\
\end{array}
\right)+\ldots\,.
\end{align}
The coefficients $x^0_{ij}(x_{ij})$ are real (complex) numbers, independent of the moduli $u_{1,3}$,
except for the choice $\bm{1}_{-++}$ at the fixed point 3, where both $x^0_{ij}$ and $x_{ij}$ are proportional to a linear combination of $u_3$ and $\bar u_3$. In any case, $x^0_{ij}(x_{ij})$ are expected to be of the same order, and we do not need to discuss separately this particular case.

Since $m_\nu(u,\bar u)$ has rank two at $x=0$, we also consider the expansion of $[m_\nu(u,\bar u)]^{-1}$, which is identical to the one in eq.~\eqref{mB}, with the replacement $m_\nu(u,\bar u)\to [m_\nu(u,\bar u)]^{-1}$ and $m_{0\nu} \to m_{0\nu}^{-1}$. The two possibilities arising from the analysis of $m_\nu(u,\bar u)$ and $[m_\nu(u,\bar u)]^{-1}$ are physically distinct. At $x=0$, the pattern of $[m_\nu(u,\bar u)]$ predicts a vanishing mass, while that of $[m_\nu(u,\bar u)]^{-1}$ predicts a
divergent neutrino mass. We understand such an infinite mass in terms of an extra degree of freedom of the full theory becoming massless at the fixed point. It is natural to interpret such a degree of freedom as a right-handed neutrino. Thus, the pattern of $[m_\nu(u,\bar u)]^{-1}$ is expected to arise in the context of the seesaw mechanism
when a right-handed neutrino, whose mass depends on the moduli, becomes massless at the fixed point.
\item[$\bullet$ {\bf C}]
\vskip 0.2 cm
\noindent
In this case, the only solution to the constraint specified in table~\ref{constraints}, is a vanishing neutrino mass matrix $m_\nu(u,\bar u)=0$, to any order in the expansion in powers of $u_1$ and $u_3$. We dismiss this unphysical possibility.
\end{itemize}
\begin{table}[h!]
\centering
%\resizebox{1.0\textwidth}{!}{
\begin{tabular}{|l|c|c|c|}
\hline\hline
{\tt pattern}&$\bm{r}_s({\tt FP 2})$&$\bm{r}_s({\tt FP 3})$&$\bm{r}_s({\tt FP 5})$\\
\hline
{\bf A}~-~eq.~\eqref{mA} & $\bm{1}_+$&$\bm{1}_{+++}$, $\bm{1}_{++-}$&$\bm{1}_+$\\
\hline
{\bf B}~-~eq.~\eqref{mB} &$\bm{1}'_-$&$\bm{1}_{-+-}$, $\bm{1}_{-++}$&$\bm{1}'_-$\\
\hline
{\bf C}&$\bm{1}_-$, &$\bm{1}_{+--}$, $\bm{1}_{+-+}$,&$\bm{1}_-$, \\
&$\bm{1}'_+$&$\bm{1}_{---}$, $\bm{1}_{--+}$&$\bm{1}'_+$\\
\hline\hline
\end{tabular}
%}
\caption{\label{numasses} Patterns of neutrino mass matrices $m_\nu(u,\bar u)$, or their inverse $m_\nu(u,\bar u)^{-1}$, and their dependence on the singlet representation of the superpotential $w$, for the fixed points $i=2,3,5$. We have displayed the singlet representations $\bm{r}_s(\texttt{FPi})$  of the stability group $G_0$. Each representation $\bm{r}_s(\texttt{FPi})$
can be embedded in one (or more) representations $\bm{r}_s$ of $N_n(H)$.}
\end{table}
From eq.~\eqref{ue} we see that moving to the basis where $m_{\bar{e}e}(u,\bar{u})$ is diagonal, up to a common permutation matrix of rows and columns and up to higher-order terms in the expansion, the neutrino mass matrix maintains the same pattern shown in eqs.~\eqref{mA} and \eqref{mB}. To first order in $x$, the effect of the basis change can be absorbed in the coefficients $x_{ij}^{0}$, $x_{ij}$. The same conclusion holds for the inverse $m_\nu(u,\bar u)^{-1}$ and, without losing generality, we can discuss the neutrino mass spectrum, mixing angles and phases by directly analyzing the matrices \eqref{mA} and \eqref{mB}. These two patterns coincide with those occurring in single-modulus $SL(2,\mathbbm{Z})$-invariant theories close to the fixed point $\tau_0=i$~\cite{Feruglio:2022koo,Feruglio:2023mii}. We briefly recapitulate the results here and we refer the reader to the literature for more details~\footnote{In particular, the predictions for the lepton masses and lepton mixing matrix as well as mixing parameters have been presented in Appendices D.1.1 and D.1.2 and Section 6 of ref.~\cite{Feruglio:2023mii}.}.

In table~\ref{synops} we summarize the predictions of Siegel modular invariant models for lepton masses in the vicinity of one of the fixed points, up to possible permutations affecting the mixing matrix. When the pattern in eq.~\eqref{mA} is realized, both normal (NO) and inverted ordering (IO) of the neutrino mass spectrum can be realized. However, $\Delta m^2_{sol}/\Delta m^2_{atm}$ is generically expected to be of order one. Moreover $\sin^2\theta_{12}$ and $\sin^2\theta_{13}$ are expected to be of the same order, contrary to observation. Therefore this pattern can only be reconciled with the data at the price of tuning the coefficients
$x_{ij}^{0}$ and $x_{ij}$.
\begin{samepage}
\begin{table}[h!]
\centering
\begin{tabular}{|lccccc|}
\hline\hline
 & $\begin{array}{c}{\rm mass}\\{\rm ordering}\end{array}$ & $\dd\frac{\Delta m^2_{sol}}{\Delta m^2_{atm}}$ & $\sin^2\theta_{12}$ & $\sin^2\theta_{13}$ & $\sin^2\theta_{23}$\\
\hline
\hline
{\bf A}~-~eq.~\eqref{mA}~~~$m_\nu(0,0)$~{\rm regular}&NO/IO&${\mathcal O}(1)$&${\mathcal O}(x^2)$&${\mathcal O}(x^2)$&${\mathcal O}(1)$\\[5 pt]
\hline
{\bf B}~-~eq.~\eqref{mB}~~~$m_\nu(0,0)$~{\rm regular}&IO&${\mathcal O}(x)$&$\frac{1}{2}(1+{\mathcal O}(x))$&${\mathcal O}(x^2)$&${\mathcal O}(1)$\\[5 pt]
\hline
{\bf B}~-~eq.~\eqref{mB}~~~$m_\nu(0,0)$~{\rm singular}&NO&${\mathcal O}(x^3)$&$\frac{1}{2}(1+{\mathcal O}(x))$&${\mathcal O}(x^2)$&${\mathcal O}(1)$\\[5 pt]
\hline
\hline
{\bf D}~-~eq.~\eqref{mD}~~~$m_\nu(0,0)$~{\rm regular}&NO/IO&${\mathcal O}(x)$&$\frac{1}{2}(1+{\mathcal O}(x))$&${\mathcal O}(x^2)$&${\mathcal O}(x^2)$\\[5 pt]
\hline
{\bf E}~-~eq.~\eqref{mE}~~~$m_\nu(0,0)$~{\rm regular}&NO/IO&${\mathcal O}(1)$&${\mathcal O}(1)$&${\mathcal O}(1)$&${\mathcal O}(1)$\\[5 pt]
\hline\hline
\end{tabular}
\caption{\label{synops} Synopsis of predictions in modular invariant flavor models of leptons, when the modulus $\tau$ falls in the vicinity of the fixed points and $\rho_L$ is an irreducible triplet. For the fixed point 4, $x=|u_1|$.}
\end{table}
\end{samepage}

When the pattern in eq.~\eqref{mB} is realized, and $m_\nu(0,0)$ has a vanishing eigenvalue, an inverted ordering of neutrino masses is predicted. To reproduce the observed values of $\sin^2\theta_{13}$ and $\sin^2\theta_{12}$, $x$ should be close to 0.15. This is in tension with the value of $x$ required by $r=\Delta m^2_{sol}/\Delta m^2_{atm}={\cal O}(x)$, experimentally close to 0.03. This pattern can match the data too, at the price of tuning the coefficients $x_{ij}^{0}$ and $x_{ij}$.

When the outcome is the pattern in eq.~\eqref{mB}, a particularly appealing scenario occurs when $[m_\nu(0,0)]^{-1}$ has a vanishing eigenvalue,
which can occur within the seesaw mechanism. In this case, all the data can be reproduced by parameters $x^0_{ij}$ and $x_{ij}$ of the same order of magnitude, by adjusting the overall scale $m_{0\nu}$ and choosing $x$ close to $0.1$. No tuning of the unknown order-one parameters is needed in this case. At the fixed point, $\Delta m^2_{sol}/\Delta m^2_{atm}=\sin\theta_{13}=\sin^2\theta_{12}-1/2=0$ and CP is
conserved. Nonvanishing values of these three quantities and CP-violating effects all originate from a small departure of $\tau$ from the fixed point. This is the most successful pattern among all those discussed in this paper.
%%%%%%%%%%%%%%%%%%%%%%%%%%%%%%%%%%%%%%%%%%%%%%%
\subsection{Fixed point 4}
%%%%%%%%%%%%%%%%%%%%%%%%%%%%%%%%%%%%%%%%%%%%%%%
From table~\ref{constraints} and the analysis of the previous Section, we find the following pattern for $m_{\bar{e}e}(u,\bar{u})$:
\begin{align}
m_{\bar ee}=m_{0e}^2~
\left(
\begin{array}{ccc}
y^0_{11}&y^{01}_{12}~ \bar u_1&y^{10}_{13}~ u_1\\
y^{01}_{12}~ u_1&y^0_{22}&y^{01}_{23} \bar u_1 \\
y^{10}_{13}~ \bar u_1 &y^{01}_{23}~ u_1&y^0_{33}
\end{array}
\right)+\ldots\,.
\end{align}
The overall real coefficient $m_{0e}^2$ has the dimension of (mass)$^2$. Dots stand for higher orders in the moduli expansion. The parameters ${y^{(0,01,10)}}_{ij}$ are real, independent of the moduli and expected to be of the same order.
To discuss the lepton mixing matrix, we diagonalize $m_{\bar e e}(u,\bar u)$:
\begin{align}
U_e^\dagger m_{\bar e e}(u,\bar u) U_e={\tt diag} [m_{\bar e e}(u,\bar u)]\,.
\end{align}
where $U_e$ is a nearly diagonal matrix:
\begin{align}
\label{ue4}
U_e=\left(
\begin{array}{ccc}
{\cal O}(1)&{\cal O}(1)~ \bar u_1&{\cal O}(1)~u_1\\
{\cal O}(1)~u_1&{\cal O}(1)&{\cal O}(1)~ \bar u_1\\
{\cal O}(1)~\bar u_1&{\cal O}(1)~u_1&{\cal O}(1)
\end{array}
\right).
\end{align}
Next, we examine the neutrino mass matrices. We find that they fall into three classes, depending on the
singlet representation $\bm{r}_s$ under which $w$ transforms.
\begin{itemize}
\item[$\bullet$ {\bf D}] Up to cyclic permutations of rows and columns, when $\bm{r}_s$ embeds
one of the $G_0$ singlets $\bm{1}_{++n}$ $(n=0,1,2)$, we get the following pattern, $m_{0\nu}$ denoting a real mass parameter:
\vskip 0.2 cm
\noindent
\begin{align}
\label{mD}
m_\nu(u,\bar u)=m_{0\nu}~
\left(
\begin{array}{ccc}
x^0_{11} &x^{01}_{12}~ \bar u_1&x^{10}_{13} ~u_1\\
\cdot&x^{10}_{22}~ u_1&x^0_{23} \\
\cdot&\cdot&x^{01}_{33}~ \bar u_1\\
\end{array}
\right)+\ldots\,.
\end{align}
The coefficients ${x^{(0,01,10)}}_{ij}$ are real and independent of the moduli $u_{1,3}$.
\item[$\bullet$ {\bf E}]
Up to cyclic permutations of rows and columns, when $\bm{r}_s$ embeds
one of the $G_0$ singlets $\bm{1}_{--n}$ $(n=0,1,2)$, we get the following pattern, $m_{0\nu}$ denoting a real mass parameter:
\vskip 0.2 cm
\noindent
\begin{align}
\label{mE}
m_\nu(u,\bar u)=m_{0\nu}~
\left(
\begin{array}{ccc}
x^{01}_{11}~ \bar u_3&x^{10}_{12}~u_3&x^{11}_{13} ~u_1\bar u_3+x^{'11}_{13} ~\bar u_1 u_3\\
\cdot&x^{11}_{22} ~u_1\bar u_3+x^{'11}_{22} ~\bar u_1 u_3& x^{01}_{23}~ \bar u_3 \\
\cdot&\cdot&x^{10}_{33}~u_3\\
\end{array}
\right)+\ldots\,.
\end{align}
The coefficients ${x^{(01,10,11)}}_{ij}$ and $x^{'11}_{ij}$ are real and independent of the moduli $u_{1,3}$.
\item[$\bullet$ {\bf F}]
When $\bm{r}_s$ embeds
one of the $G_0$ singlets $\bm{1}_{+-n}$ or $\bm{1}_{-+n}$ $(n=0,1,2)$, to all orders in the moduli expansion
the neutrino mass matrix vanishes, an option we discard as unphysical.

Given the form of $U_e$ in eq.~\eqref{ue4}, in the basis where $m_{\bar{e}e}(u,\bar{u})$ is diagonal, up to a common permutation matrix of rows and columns and up to higher-order terms in the expansion, the neutrino mass matrix has always the same pattern shown in eqs.~\eqref{mD} and \eqref{mE}. To first order in $u_1$ and $\bar u_1$, the effect of the basis change can be absorbed in the coefficients ${x^{(01,10,11)}}_{ij}$ and $x^{'11}_{ij}$. Thus, without losing generality, we can discuss the neutrino mass spectrum, mixing angles and phases by directly analyzing the matrices \eqref{mD} and \eqref{mE}.
\end{itemize}
The pattern in eq.~\eqref{mD} coincides with the one occurring in single-modulus $SL(2,\mathbbm{Z})$-invariant theories close to the fixed point $\tau_0=\omega$~\cite{Feruglio:2023mii}. We refer the reader to the literature for more details~\footnote{In particular, the predictions for the lepton masses and lepton mixing matrix as well as mixing parameters have been presented in Appendix D.2 and Section 6 of ref.~\cite{Feruglio:2023mii}.}.
In table~\ref{synops} we summarize the predictions of Siegel modular invariant models for lepton masses in the vicinity of the fixed point 4, up to possible permutations affecting the mixing matrix. Within the pattern in eq.~\eqref{mD}, both normal and inverted ordering of the neutrino mass spectrum can be realized. However, $\Delta m^2_{sol}/\Delta m^2_{atm}={\cal O}(x)$ would demand $x\approx 0.03$, which is inadequate to describe $\sin^2\theta_{13}$ and $\sin^2\theta_{23}$, both expected of ${\cal O}(x^2)$. A considerable tuning of the order-one coefficients is required to reconcile this pattern with the data. Finally, the pattern in eq.~\eqref{mE} predicts neutrino masses of the same order of magnitude and mixing angle of
approximately the same size, as a numerical simulation shows. While data can be reproduced by fitting the unknown order-one coefficients, this pattern does not suggest any explanation for the smallness of $\Delta m^2_{sol}/\Delta m^2_{atm}$ and/or $\sin^2\theta_{13}$.

%%%%%%%%%%%%%%%%%%%%%%%%%%%%%%%%%%%%%%%%%%%%%%%
\subsection{Summary}
%%%%%%%%%%%%%%%%%%%%%%%%%%%%%%%%%%%%%%%%%%%%%%%
Excluding the unphysical case of a vanishing neutrino mass matrix, the four patterns of eqs.~(\ref{mA},\ref{mB},\ref{mD},\ref{mE}) exhaust all possible cases that can arise from a CP and Siegel modular invariant locally supersymmetric theory when the moduli are close to a fixed point.
The predictions are summarized in table~\ref{synops}, where the scaling properties of the leptonic mixing angles and the ratio between solar and atmospheric squared mass differences are shown. The distance from the fixed point is parametrized by a small variable $x$. The only pattern that accommodates all the data without tuning of the unknown parameters is {\bf B} of eq.~\eqref{mB}, when $m_\nu(0,0)$ is singular. It can be realized by the seesaw mechanism when one of the right-handed neutrinos happens to be massless at the fixed point. All the other patterns require an adjustment of the parameters to overcome the wrong scaling of one or several observable quantities. An explicit model belonging to the class analyzed in this Section has been discussed in ref.~\cite{Ding:2021iqp} and, for completeness, is illustrated in the Appendix~\ref{amodel}, where we show that it matches pattern {\bf A} of eq.~\eqref{mA}.

Many properties of the mass spectrum and the mixing matrix follow mainly from the decomposition of the representation $\Omega$ into irreducible components. This idea was developed in ref.~\cite{Reyimuaji:2018xvs}, where the decompositions of $\Omega$, and its charged lepton counterpart $\Omega_c$, compatible with a realistic leading-order pattern of lepton masses and mixing angles have been classified.
While our results apply to a specific context and are not aimed to cover the case of the most general flavour group acting linearly on matter fields, they have been obtained under less restrictive assumptions.
First, we go beyond the zeroth order approximation by including the correction linear in the moduli.
Second, we also consider the case where the neutrino mass matrix develops diverging eigenvalues at the
fixed point, as happens in the seesaw mechanism when a right-handed neutrino mass
vanishes at the symmetric point. Third, consistently with the freedom permitted by supergravity, we
allow an overall phase factor in the transformation of neutrino mass matrices.
%%%%%%%%%%%%%%%%%%%%%%%%%%%%%%%%%%%%%%%%%%%%%%%
\section{Conclusion and outlook}
%%%%%%%%%%%%%%%%%%%%%%%%%%%%%%%%%%%%%%%%%%%%%%%
To explore theories depending on more than one modulus, we have analyzed a class of locally supersymmetric models of lepton masses invariant under CP and under a subgroup $N(H)$ of the Siegel modular group $Sp(4,\mathbbm{Z})$. We have concentrated on the subgroup $N(H)$ whose finite copies $N_n(H)$ contain three-dimensional irreducible representations, to which we assign lepton electroweak doublets. For consistency, the moduli space is restricted to a subset $\Sigma_2$ of the Siegel upper half plane, spanned by two complex moduli and invariant under $N(H)$. We have identified 168 irreducible triplets of $N_n(H)$, which we conjecture to exhaust the number of inequivalent three-dimensional irreducible representations. There are four inequivalent fixed points in $\Sigma_2$, each left invariant by a specific finite group $G_0$. By exploiting the decomposition of each irreducible
triplet of $N_n(H)$ under $G_0$, and a convenient basis for moduli and matter fields, we have built all possible patterns of neutrino mass matrices, consisting of a series expansion around each fixed point, of which we keep the first nontrivial term. The leading-order contribution is invariant under both $G_0$ and CP, which are spontaneously broken by the 
vacuum expectation value of an order parameter, measuring the distance from the fixed point.
After moving to the basis where kinetic terms are canonical and the charged lepton mass matrix is diagonal, we can read neutrino masses and lepton mixing angles. 

Apart from the unrealistic case of a vanishing neutrino mass matrix, only five patterns are found. Four of them coincide with those arising
in $SL(2,\mathbbm{Z})$-invariant single-modulus theories in the vicinity of the fixed points $\tau_0=i$ and $\tau_0=-1/2+i~\sqrt{3}/2$. In each pattern, all physical quantities scale with the distance of the moduli from the fixed point in a way that is largely independent of the details of the theory.
All the patterns but a single one require tuning the free parameters to match such a scaling with the smallness of $\Delta m^2_{sol}/\Delta m^2_{atm}$ and $\sin^2\theta_{13}$ and the largeness of $\sin^2\theta_{12}$ and $\sin^2\theta_{23}$. The matrix that best describes the data and needs no tuning is pattern {\bf B} of eq.~\eqref{mB}, when $m_\nu(0,0)$ is singular.
This is the same pattern preferred by the majority of single-modulus models when $\tau$ is close to the self-dual point $\tau_0=i$.

We stress the generality of this result. Except for the assumption that the lepton doublets transform under $N_n(H)$ as any of the 168 irreducible triplets, our finding is independent of the level, of the weights
of matter multiplets, and the form of the K\"ahler potential. In particular, we are not forced to assume a minimal or flavor universal K\"ahler potential:
our conclusion holds for the most general K\"ahler potential compatible with Siegel modular invariance. It would be very difficult and time-consuming to reproduce our results by inspecting one-by-one all models compatible with our assumptions. Today very few models of this type exist in the literature~\cite{Ding:2020zxw,Ding:2021iqp,Kikuchi:2023dow,RickyDevi:2024ijc}. When moduli fall in the vicinity of a fixed point, we find agreement with the present analysis.

On the other hand, our analysis has some limitations. First, the dependence on the chosen level, weights, and K\"ahler potential affect the unknown coefficients of the patterns we found. We have assumed that such dependence does not conspire to produce hierarchies among these coefficients, but we cannot provide a mathematical proof of this statement, which seems to work well in the case of single-modulus theories. Second, we have no dynamical justification for working close to a fixed point. Here again, we rely on the statistics accumulated in the single modulus theories and on the few semianalytical results derived from the minimization of modular invariant energy densities. Finally, new patterns may arise if we relax the assumption
that the lepton doublets transform as irreducible triplets of the finite modular group. This is a realistic possibility, especially in the context of string theory compactifications where typically the modular group is part of a bigger eclectic group~\cite{Baur:2019kwi,Nilles:2020nnc,Nilles:2020kgo,Nilles:2020tdp,Baur:2020jwc,Nilles:2020gvu,Baur:2021mtl}. Allowing for reducible representations would presumably open many new unexplored possibilities. Both the two-dimensional region $\Sigma_1$ or the entire Siegel upper half plane might be relevant in a more general analysis. A complete classification of all representations of dimensionality one and two would be required to undertake such a demanding task, which we leave for future work.
\vskip 0.5 cm
\noindent
%%%%%%%%%%%%%%%%%%%%%%%%%%%%%%%%%%%%%%%
\section*{Acknowledgements}
We thank warmly Gianguido Dall'Agata and Davide Cassani, for a stimulating correspondence. This work is supported by the INFN. GJD is supported by the National Natural Science Foundation of China under Grant Nos. 12375104 and 11975224. XGL is supported by the National Science Foundation, under Grant No.\ PHY-1915005.
%%%%%%%%%%%%%%%%%%%%%%%%%%%%%%%%%%%%%%%
\newpage
%%%%%%%%%%%%%%%%%%%%%%%%%%%%%%%%%%%%%%%%%%%%%%%
\setcounter{equation}{0}
\renewcommand{\theequation}{\thesection.\arabic{equation}}
\begin{appendix}
%%%%%%%%%%%%%%%%%%%%%%%%%%%%%%%%%%%%%%%%%%%%%%%
\section{\label{app:groupD4}Group theory of $D_4$}
%%%%%%%%%%%%%%%%%%%%%%%%%%%%%%%%%%%%%%%%%%%%%%%
The dihedral group $D_4$ is the symmetry group of a square, generated by the two elements $a$ and $b$, satisfying $a^4=b^2=(ab)^2=1$. The element $a$ represents a rotation of $\pi/2$ of the square around its center, while the element $b$ is a reflection around a symmetry axis. The group has
5 conjugacy classes and 5 irreducible representations: four singlets $\bm{1}_+$, $\bm{1}_-$, $\bm{1}'_+$, $\bm{1}'_-$ and 1 doublet $\bm{2}$. The representation matrices of the generators $a$ and $b$ are
\begin{align}
\label{eq:D4-irrep}
\begin{array}{ccc}
\bm{1}_+:~~&~a=+1,~&~b=+1\,, \\
\bm{1}_-:~~&~a=+1,~&~b=-1\,,\\
\bm{1}'_-:~~&~a=-1,~&~b=+1\,,\\
\bm{1}'_+:~~&~a=-1,~&~b=-1\,,\\
\bm{2}:~~&~a=\left(\begin{array}{cc}
0 & -1 \\
1 & 0 \end{array}\right),~&~b=\left(\begin{array}{cc}
1 & 0 \\
0 & -1
\end{array}\right)\,.
\end{array}
\end{align}
\begin{table}[h]
\begin{center}
\begin{tabular}{|c|c|c|c|c|c|}
\hline\hline
 & $1C_1$  & $1C_2$ & $2C_2$ & $2C'_2$ & $2C_4$\\
\hline
&$\{b^2\}$ &$\{a^2\}$&$\{b,a^2b\}$&$\{ab, a^3b\}$  &$\{a,a^3\}$\\
\hline
$\bm{1}_+$ &$+1$ & $+1$ & $+1$ & $+1$ &$+1$\\
\hline
$\bm{1}_-$ & $+1$ & $+1$ & $-1$ & $-1$ & $+1$\\
\hline
$\bm{1}'_-$ & $+1$ & $+1$ & $+1$ & $-1$ & $-1$\\
\hline
$\bm{1}'_+$ & $+1$ & $+1$ & $-1$ & $+1$ & $-1$ \\
\hline
$\bm{2}$ & $+2$ & $-2$ & $0$ & 0 & $0$\\
\hline\hline
\end{tabular}
\end{center}
\caption{Character table of $D_4$, where $mC_n$
denotes a conjugacy class with $m$ elements of order $n$.}
\label{cD4}
\end{table}%

\noindent
The character table is shown in table~\ref{cD4}. The product of representations decomposes as follows:
\begin{align}
&\bm{1}_{m}\otimes \bm{1}_{n}=\bm{1}_{mn},\qquad \bm{1}_{m}\otimes \bm{1^\prime}_{n}=\bm{1^{\prime}}_{mn} , \qquad  \bm{1^\prime}_{m}\otimes \bm{1^\prime}_{n}=\bm{1}_{mn}\nn\\
&\bm{1}_{m}\otimes \bm{2}=\bm{2},\qquad \bm{1}'_{m}\otimes \bm{2}=\bm{2}\,,\qquad \bm{2}\otimes \bm{2}= \bm{1}_{+}\oplus \bm{1}_{-}\oplus \bm{1^\prime}_{+}\oplus \bm{1^\prime}_{-}\,.
\end{align}
\noindent
We define:
\begin{align}\nn
\bm{1}_{\pm}=\beta_{\pm}\quad\quad\quad \bm{1'}_{\pm}=\beta'_{\pm}\quad\quad\quad
\bm{2}=
\left(
\begin{array}{c}
\alpha_1\\
\alpha_2
\end{array}
\right)
\end{align}
and we get:
\begin{align}
\bm{1}_{+}\otimes \bm{2}=\bm{2}=
\left(
\begin{array}{c}
\beta_+ \alpha_1\\
\beta_+ \alpha_2
\end{array}
\right)\,,\qquad &\bm{1}_{-}\otimes \bm{2}=\bm{2}=
\left(
\begin{array}{c}
\beta_- \alpha_2\\
-\beta_- \alpha_1
\end{array}
\right)\,.\\
\bm{1'}_{+}\otimes \bm{2}=\bm{2}=
\left(
\begin{array}{c}
\beta'_+ \alpha_2\\
\beta'_+ \alpha_1
\end{array}
\right)\,,\qquad &\bm{1'}_{-}\otimes \bm{2}=\bm{2}=
\left(
\begin{array}{c}
\beta'_- \alpha_1\\
-\beta'_- \alpha_2
\end{array}
\right)\,.
\end{align}

%\begin{align}\nn
%\bm{2}\otimes \bm{2}= \bm{1}_{+}\oplus \bm{1}_{-}\oplus \bm{1^\prime}_{+}\oplus \bm{1^\prime}_{-},
%~~~~~~~~~~~~~~~~~~~~~~
%\bm{1}_{+}=&~\alpha_1\beta_1+\alpha_2\beta_2\nn\\
%\bm{1}_{-}=&~\alpha_1\beta_2-\alpha_2\beta_1\nn\\
%\bm{1}'_{-}=&~\alpha_1\beta_1-\alpha_2\beta_2\nn\\
%\bm{1}'_{+}=&~\alpha_1\beta_2+\alpha_2\beta_1\nn.
%\end{align}
\begin{align}
\bm{2}\otimes \bm{2}= \bm{1}_{+}\oplus \bm{1}_{-}\oplus \bm{1^\prime}_{+}\oplus \bm{1^\prime}_{-},~~~~~~~~\left\{
\begin{array}{c}
\bm{1}_{+}=~\alpha_1\beta_1+\alpha_2\beta_2  \\
\bm{1}_{-}=~\alpha_1\beta_2-\alpha_2\beta_1 \\
\bm{1}'_{-}=~\alpha_1\beta_1-\alpha_2\beta_2\\
\bm{1}'_{+}=~\alpha_1\beta_2+\alpha_2\beta_1
\end{array}\right.\,.
\end{align}
An equivalent basis $(a',b')$ for the $D_4$ generators is obtained by the similarity transformation acting on doublets:
\be
a'=U^{-1} a~ U=
\left(\begin{array}{cc}
i & 0 \\
0 & -i\end{array}\right)\,,~~~~~~~~b'=U^{-1} b~ U=
\left(\begin{array}{cc}
0 & 1 \\
1 & 0 \end{array}\right),
\ee
where $U$ is the matrix
\be
U=\frac{1}{\sqrt{2}}
\left(
\begin{array}{cc}
i&i\\
1&-1
\end{array}
\right).
\ee
Singlet representations are unchanged in this new basis, where the consistency conditions of CP transformation in eq.~\eqref{eq:gCP-cons-FP5} are solved by $X_I'=\mathbbm{1}$, both for singlets and doublets when working at the fixed point 5. Of course, this new basis is inconvenient when dealing with the fixed point 2, since a nontrivial matrix
$X_I'$ would be required in the CP transformation law of doublets.
\newpage
\setcounter{equation}{0}
%%%%%%%%%%%%%%%%%%%%%%%%%%%%%%%%%%%%%%%%%%%%%%%
\section{\label{app:groupPauli}Group theory of $D_4\circ Z_4$}
%%%%%%%%%%%%%%%%%%%%%%%%%%%%%%%%%%%%%%%%%%%%%%%
The Pauli group is the central product of $D_4$ and $Z_4$: $D_4\circ Z_4$, which has the GAP id [16,13] and is isomorphic to $(Z_2\times Z_4)\rtimes Z_2$. It is generated by the elements $a$, $b$ and $c$, satisfying $a^4=c^4=b^2=(ab)^2=1$, $a^2=c^2$, $ac=ca$, $bc=cb$. The elements $(a,b)$ generate $D_4$ and $c$ generates $Z_4$. The elements of $D_4$ commute with those of $Z_4$. They have in common the element $a^2=c^2$ and for this reason the central product does not coincide with the direct product. The group has
ten irreducible representations: eight singlets and two doublets. The generators $a$, $b$ and $c$ in each irreducible representation are represented by
\begin{align}
\label{eq:D4oZ4-irrep}
\begin{array}{cccc}
\bm{1}_{+++}:~~&~a=+1,~&~b=+1,~&~c=+1\,,\\
\bm{1}_{+--}:~~&~a=+1,~&~b=-1,~&~c=-1\,, \\
\bm{1}_{+-+}:~~&~a=+1,~&~b=-1,~&~c=+1\,,\\
\bm{1}_{---}:~~&~a=-1,~&~b=-1,~&~c=-1\,, \\
\bm{1}_{--+}:~~&~a=-1,~&~b=-1,~&~c=+1\,, \\
\bm{1}_{-+-}:~~&~a=-1,~&~b=+1,~&~c=-1\,, \\
\bm{1}_{-++}:~~&~a=-1,~&~b=+1,~&~c=+1\,, \\
\bm{1}_{++-}:~~&~a=+1,~&~b=+1,~&~c=-1\,, \\
\bm{2}:~~&~a=\left(\begin{array}{cc}
i & 0 \\
0 & -i \end{array}\right),~&~b=\left(\begin{array}{cc}
0 & 1 \\
1 & 0 \end{array}\right),~&~c=\left(\begin{array}{cc}
-i & 0 \\
0 & -i
\end{array}\right)\,, \\ [0.3in]
\bm{2}':~~&~a=\left(\begin{array}{cc}
i & 0 \\
0 & -i \end{array}\right),~&~b=\left(\begin{array}{cc}
0 & 1 \\
1 & 0
\end{array}\right),~&~c=\left(\begin{array}{cc}
i & 0 \\
0 & i
\end{array}\right)\,.
\end{array}
\end{align}
\begin{table}[h]
\begin{center}
\scalebox{0.85}{
\begin{tabular}{|c|c|c|c|c|c|c|c|c|c|c|}
\hline\hline
%&$1a$ & $2d$ & $4e$ & $4b$ &$2b$ & $2c$ & $2a$ &$4c$&$4d$ &$4a$\\\hline
& $1C_1$ & $1C_2$ & $1C_4$  & $1C'_4$ & $2C_2$  & $2C'_2$ & $2C''_2$   & $2C_4$ & $2C'_4$ & $2C''_4$ \\\hline
&$\{b^2\}$ & $\{c^2\}$ &$\{c\}$ & $\{a^2c\}$ &$\{b, a^2b\}$ &$\{ab, abc^2\}$ & $\{ac, ac^3\}$  &$\{a, ac^2\}$&$\{bc, a^2bc\}$ &$\{abc, abc^3\}$\\
\hline
$\bm{1}_{+++}$ &$+1$ &$+1$ & $+1$&$+1$&$+1$&$+1$&$+1$&$+1$&$+1$&$+1$\\
\hline
$\bm{1}_{+--}$ & $+1$ & $+1$ & $-1$ & $-1$ & $-1$ & $-1$ & $-1$ & $+1$ & $+1$ & $+1$ \\ \hline
$\bm{1}_{+-+}$ & $+1$ & $+1$ & $+1$ & $+1$ & $-1$ & $-1$ & $+1$ & $+1$ & $-1$ & $-1$\\ \hline
$\bm{1}_{---}$ & $+1$ & $+1$ & $-1$ & $-1$ & $-1$ & $+1$ & $+1$ & $-1$ & $+1$ & $-1$ \\ \hline
$\bm{1}_{--+}$ & $+1$ & $+1$ & $+1$ & $+1$ & $-1$ & $+1$ & $-1$ & $-1$ & $-1$ & $+1$ \\\hline
$\bm{1}_{-+-}$ & $+1$ & $+1$ & $-1$ & $-1$ & $+1$ & $-1$ & $+1$ & $-1$ & $-1$ & $+1$ \\ \hline
$\bm{1}_{-++}$ & $+1$ & $+1$ & $+1$ & $+1$ & $+1$ & $-1$ & $-1$ & $-1$ & $+1$ & $-1$ \\ \hline
$\bm{1}_{++-}$ & $+1$ & $+1$ & $-1$ & $-1$ & $+1$ & $+1$ & $-1$ & $+1$ & $-1$ &$-1$ \\ \hline
$\bm{2}$ & $+2$ & $-2$ & $-2i$ & $+2i$ & $0$ & $0$ & $0$ & $0$ & $0$ & $0$\\ \hline
$\bm{2}'$ & $+2$ & $-2$ & $+2i$ & $-2i$ & $0$ & $0$ & $0$ & $0$ & $0$ & $0$\\
\hline\hline
\end{tabular}}
\end{center}
\caption{Character table of $D_4\circ Z_4$.}
\label{cPauli}
\end{table}%
The character table is shown in table~\ref{cPauli}. The product of representations decomposes as follows:
%\be
%\bm{1}_{mnp}\otimes\bm{1}_{m'n'p'}=\bm{1}_{(mm')(nn')(pp')}\,.
%\ee
\begin{align}
\bm{1}_{mnp}\otimes\bm{1}_{m'n'p'}=&\bm{1}_{(mm')(nn')(pp')}\,,\nn\\
\bm{2}\otimes\bm{2}=\bm{2}'\otimes\bm{2}'=&~\bm{1}_{---}\oplus\bm{1}_{+--}\oplus\bm{1}_{-+-}\oplus\bm{1}_{++-}\,,\nn\\
\bm{2}\otimes\bm{2}'=&~\bm{1}_{+++}\oplus\bm{1}_{-++}\oplus\bm{1}_{+-+}\oplus\bm{1}_{--+}\,,\nn\\
&\hskip-1.2in\left(\begin{array}{c}\bm{2}\\\bm{2}'\end{array}\right)\otimes
\left(\bm{1}_{+++},\bm{1}_{-++},\bm{1}_{+-+},\bm{1}_{--+}\right)=~
\left(\begin{array}{c}\bm{2}\\\bm{2}'\end{array}\right)\,,\nn\\
&\hskip-1.2in \left(\begin{array}{c}\bm{2}\\\bm{2}'\end{array}\right)\otimes
\left(\bm{1}_{---},\bm{1}_{+--},\bm{1}_{-+-},\bm{1}_{++-}\right)=~
\left(\begin{array}{c}\bm{2}'\\\bm{2}\end{array}\right)\,.
\end{align}
%\begin{align}
%\left(\begin{array}{c}\bm{2}\\\bm{2}'\end{array}\right)\otimes
%\left(\bm{1}_{+++},\bm{1}_{-++},\bm{1}_{+-+},\bm{1}_{--+}\right)=&~
%\left(\begin{array}{c}\bm{2}\\\bm{2}'\end{array}\right)\,,\nn\\
%\left(\begin{array}{c}\bm{2}\\\bm{2}'\end{array}\right)\otimes
%\left(\bm{1}_{---},\bm{1}_{+--},\bm{1}_{-+-},\bm{1}_{++-}\right)=&~
%\left(\begin{array}{c}\bm{2}'\\\bm{2}\end{array}\right)\,.
%\end{align}

\noindent
The tensor products between singlets and doublets are given by
\begin{align}
\nn
\bm{1}_{+++}\otimes \bm{2}=\bm{2}=
\alpha\begin{pmatrix}
\beta_1\\
\beta_2
\end{pmatrix}\,,
&&\bm{1}_{+++}\otimes \bm{2}'=\bm{2}'=
\alpha\begin{pmatrix}
\beta_1\\
\beta_2
\end{pmatrix}\,, \\
\nn \bm{1}_{++-}\otimes \bm{2}=\bm{2}'=
\alpha\begin{pmatrix}
\beta_1\\
\beta_2
\end{pmatrix}\,,
&&\bm{1}_{++-}\otimes \bm{2}'=\bm{2}=
\alpha\begin{pmatrix}
\beta_1\\
\beta_2
\end{pmatrix}\,,\\
\nn
\bm{1}_{+--}\otimes \bm{2}=\bm{2}'=
\alpha\begin{pmatrix}
\beta_1\\
-\beta_2
\end{pmatrix}\,,
&&\bm{1}_{+--}\otimes \bm{2}'=\bm{2}=
\alpha\begin{pmatrix}
\beta_1\\
-\beta_2
\end{pmatrix}\,, \\
\nn
\bm{1}_{+-+}\otimes \bm{2}=\bm{2}=
\alpha\begin{pmatrix}
\beta_1\\
-\beta_2
\end{pmatrix}\,,
&&\bm{1}_{+-+}\otimes \bm{2}'=\bm{2}'=
\alpha\begin{pmatrix}
\beta_1\\
-\beta_2
\end{pmatrix}\,, \\
\nn
\bm{1}_{---}\otimes \bm{2}=\bm{2}'=
\alpha\begin{pmatrix}
\beta_2\\
-\beta_1
\end{pmatrix}\,,
&&\bm{1}_{---}\otimes \bm{2}'=\bm{2}=
\alpha\begin{pmatrix}
\beta_2\\
-\beta_1
\end{pmatrix}\,, \\
\nn
\bm{1}_{--+}\otimes \bm{2}=\bm{2}=
\alpha\begin{pmatrix}
\beta_2\\
-\beta_1
\end{pmatrix}\,,
&&\bm{1}_{--+}\otimes \bm{2}'=\bm{2}'=
\alpha\begin{pmatrix}
\beta_2\\
-\beta_1
\end{pmatrix}\,, \\
\nn
\bm{1}_{-+-}\otimes \bm{2}=\bm{2}'=
\alpha\begin{pmatrix}
\beta_2\\
\beta_1
\end{pmatrix}\,,
&&\bm{1}_{-+-}\otimes \bm{2}'=\bm{2}=
\alpha\begin{pmatrix}
\beta_2\\
\beta_1
\end{pmatrix}\,, \\
\bm{1}_{-++}\otimes \bm{2}=\bm{2}=
\alpha\begin{pmatrix}
\beta_2\\
\beta_1
\end{pmatrix}\,,
&&\bm{1}_{-++}\otimes \bm{2}'=\bm{2}'=
\alpha\begin{pmatrix}
\beta_2\\
\beta_1
\end{pmatrix}\,.
\end{align}
The tensor products between two doublets are
\begin{align}\nn
\bm{2}\otimes \bm{2}= \bm{1}_{+--}\oplus \bm{1}_{---}\oplus \bm{1}_{-+-}\oplus \bm{1}_{++-},~~~~~~~~\left\{
\begin{array}{c}
\bm{1}_{+--}=~\alpha_1\beta_2-\alpha_2\beta_1  \\
\bm{1}_{---}=~\alpha_1\beta_1-\alpha_2\beta_2 \\
\bm{1}_{-+-}=~\alpha_1\beta_1+\alpha_2\beta_2\\
\bm{1}_{++-}=~\alpha_1\beta_2+\alpha_2\beta_1
\end{array}\right.\,,\\
\nn \bm{2}\otimes \bm{2}'= \bm{1}_{+++}\oplus \bm{1}_{+-+}\oplus \bm{1}_{--+}\oplus \bm{1}_{-++},~~~~~~~~\left\{
\begin{array}{c}
\bm{1}_{+++}=~\alpha_1\beta_2+\alpha_2\beta_1  \\
\bm{1}_{+-+}=~\alpha_1\beta_2-\alpha_2\beta_1 \\
\bm{1}_{--+}=~\alpha_1\beta_1-\alpha_2\beta_2\\
\bm{1}_{-++}=~\alpha_1\beta_1+\alpha_2\beta_2
\end{array}\right.\,,\\
\bm{2}'\otimes \bm{2}'= \bm{1}_{+--}\oplus \bm{1}_{---}\oplus \bm{1}_{-+-}\oplus \bm{1}_{++-},~~~~~~~~\left\{
\begin{array}{c}
\bm{1}_{+--}=~\alpha_1\beta_2-\alpha_2\beta_1  \\
\bm{1}_{---}=~\alpha_1\beta_1-\alpha_2\beta_2 \\
\bm{1}_{-+-}=~\alpha_1\beta_1+\alpha_2\beta_2\\
\bm{1}_{++-}=~\alpha_1\beta_2+\alpha_2\beta_1
\end{array}\right.\,,
\end{align}
\newpage
\setcounter{equation}{0}
%%%%%%%%%%%%%%%%%%%%%%%%%%%%%%%%%%%%%%%%%%%%%%%
\section{\label{app:groupD4Z3}Group theory of $D_4\times Z_3$}
%%%%%%%%%%%%%%%%%%%%%%%%%%%%%%%%%%%%%%%%%%%%%%%
The group $D_4\times Z_3$ is generated by the elements $a$ and $b$ and $c$, satisfying $a^4=b^2=(ab)^2=c^3=1$, $ca=ac$, $cb=bc$.
The elements $(a,b)$ generate $D_4$ and $c$ generates $Z_3$.
This group has 15 conjugacy classes as follows,
\begin{eqnarray}
\nonumber&&1C_1=\left\{1 \right\}\,,\\
\nonumber&&1C_2=\left\{a^2 \right\}\,,\\
\nonumber&&1C_3=\left\{c \right\}\,,\\
\nonumber&&1C'_3=\left\{c^2 \right\}\,,\\
\nonumber&&1C_6=\left\{a^2 c \right\}\,,\\
\nonumber&&1C'_6=\left\{a^2c^2 \right\}\,,\\
\nonumber&&2C_2=\left\{b, a^2b \right\}\,,\\
\nonumber&&2C'_2=\left\{ab, a^3b \right\}\,,\\
\nonumber&&2C_6=\left\{bc, a^2bc \right\}\,,\\
\nonumber&&2C'_6=\left\{bc^2, a^2bc^2 \right\}\,,\\
\nonumber&&2C''_6=\left\{abc, a^3bc \right\}\,,\\
\nonumber&&2C^{(3)}_6=\left\{abc^2, a^3bc^2 \right\}\,,\\
\nonumber&&2C_4=\left\{a, a^3 \right\}\,,\\
\nonumber&&2C_{12}=\left\{ac, a^3c \right\}\,,\\
&&2C'_{12}=\left\{ac^2, a^3c^2 \right\}\,.
\end{eqnarray}
%The stability group $D_4\times Z_3$ has twelve singlet representations and %three doublet representations,
The irreducible representations of $D_4\times Z_3$ are 12 singlets and 3 doublets, the representation matrices of the generators are
%\begin{eqnarray}
\begin{align}
\label{eq:D4xZ3-irrep} \begin{array}{cccc}
\bm{1}_{++n}:~~&~a=+1,~&~b=+1,~&~c=\omega^n\,, \\
\bm{1}_{+-n}:~~&~a=+1,~&~b=-1,~&~c=\omega^n\,, \\
\bm{1}_{-+n}:~~&~a=-1,~&~b=+1,~&~c=\omega^n\,, \\
\bm{1}_{--n}:~~&~a=-1,~&~b=-1,~&~c=\omega^n\,,\\
\bm{2}_n:~~&~a=\left(\begin{array}{cc}+i&0\\0&-i\end{array}\right),~&~b=\left(\begin{array}{cc}0&1\\1&0\end{array}\right),
~&~c=\left(\begin{array}{cc} \omega^n & 0\\
0 & \omega^n
\end{array}\right)\,,
\end{array}
\end{align}
%\end{eqnarray}
with $n=0, 1, 2$ and $\omega=e^{2\pi i/3}$.
The character table is listed in table~\ref{tab:cha-tab-D4xZ3-FP4}.
\begin{table}[h]
\begin{center}
\scalebox{0.85}{
%\resizebox{1.0\textwidth}{!}{
\begin{tabular}{|c|c|c|c|c|c|c|c|c|c|c|c|c|c|c|c|}
\hline\hline
 & $1C_1$ & $1C_2$  & $1C_3$  & $1C'_3$  & $1C_6$  & $1C'_6$ & $2C_2$  & $2C'_2$ & $2C_6$  & $2C'_6$ & $2C''_6$ &
  $2C^{(3)}_6$ & $2C_4$ & $2C_{12}$ & $2C'_{12}$ \\ \hline
%&$\{b^2\}$ & $\{a^2\}$ & $\{c\}$ & $\{c^2\}$ & $\{a^2c\}$ & $\{a^2c^2\}$ & $\{b, a^2b\}$ & $\{ab, a^3b\}$  &  $\{bc, a^2bc\}$ &  $\{bc^2, a^2bc^2\}$ &  $\{abc, a^3bc\}$ &  $\{abc^2, a^3bc^2\}$ & $\{a, a^3\}$  & $\{ac, a^3c\}$ & $\{ac^2, a^3c^2\}$ \\ \hline
% &$\{b^2\}$ & $\{a^2\}$ & $\{c\}$ & $\{c^2\}$ & $\{a^2c\}$ & $\{a^2c^2\}$ &$\begin{array}{l}
%\{b,\\
% ~~a^2b\}
%\end{array}$ & $\begin{array}{l}
%\{ab,\\
%~~ a^3b\}
%\end{array}$  &  $\begin{array}{l}
%\{bc,\\
%~~a^2bc\}
%\end{array}$ &  $\begin{array}{l}
%\{bc^2,\\
%~~a^2bc^2\}
%\end{array}$ &  $\begin{array}{l}
%\{abc,\\
%~~a^3bc\}
%\end{array}$ &  $\begin{array}{l}
%\{abc^2, \\
%~~a^3bc^2\}
%\end{array}$ & $\begin{array}{l}
%\{a,\\
%~~a^3\}
%\end{array}$  & $\begin{array}{l}
%\{ac,\\
%~~a^3c\}
%\end{array}$ & $\begin{array}{l}
%\{ac^2,\\
%~~a^3c^2\}
%\end{array}$ \\ \hline
$\bm{1}_{++0}$ & 1 & 1 & 1 & 1 & 1 & 1 & 1 & 1 & 1 & 1 & 1 & 1 & 1 & 1 & 1 \\ \hline
$\bm{1}_{++1}$ & 1 & 1 & $\omega$  & $\omega^2$ & $\omega$  & $\omega^2$ & 1 & 1 & $\omega$  & $\omega^2$ &  $\omega$  & $\omega^2$ & 1 & $\omega$  & $\omega^2$ \\ \hline
$\bm{1}_{++2}$ & 1 & 1 & $\omega^2$ & $\omega$  & $\omega^2$ & $\omega$  & 1 & 1 & $\omega^2$ & $\omega$  &  $\omega^2$ & $\omega$  & 1 & $\omega^2$ & $\omega$  \\ \hline
$\bm{1}_{+-0}$ & 1 & 1 & 1 & 1 & 1 & 1 & $-1$ & $-1$ & $-1$ & $-1$ & $-1$ & $-1$ & 1 & 1 & 1 \\ \hline
$\bm{1}_{+-1}$ & 1 & 1 & $\omega$  & $\omega^2$ & $\omega$  & $\omega^2$ & $-1$ & $-1$ & $-\omega$  & $-\omega^2$ & $-\omega$  & $-\omega^2$ & 1 & $\omega$  & $\omega^2$ \\ \hline
$\bm{1}_{+-2}$ & 1 & 1 & $\omega^2$ & $\omega$  & $\omega^2$ & $\omega$  & $-1$ & $-1$ & $-\omega^2$ & $-\omega$  & $-\omega^2$ & $-\omega$  & 1 & $\omega^2$ & $\omega$  \\ \hline
$\bm{1}_{-+0}$ & 1 & 1 & 1 & 1 & 1 & 1 & 1 & $-1$ & 1 & 1 & $-1$ & $-1$ & $-1$ & $-1$ & $-1$ \\ \hline
$\bm{1}_{-+1}$ & 1 & 1 & $\omega$  & $\omega^2$ & $\omega$  & $\omega^2$ & 1 & $-1$ & $\omega$  & $\omega^2$ & $-\omega$  & $-\omega^2$ & $-1$ & $-\omega$  & $-\omega^2$ \\ \hline
$\bm{1}_{-+2}$ & 1 & 1 & $\omega^2$ & $\omega$  & $\omega^2$ & $\omega$  & 1 & $-1$ & $\omega^2$ & $\omega$  & $-\omega^2$   & $-\omega$  & $-1$ & $-\omega^2$ & $-\omega$  \\ \hline
$\bm{1}_{--0}$ & 1 & 1 & 1 & 1 & 1 & 1 & $-1$ & 1 & $-1$ & $-1$ & 1 & 1 & $-1$ & $-1$ & $-1$ \\ \hline
$\bm{1}_{--1}$ & 1 & 1 & $\omega$  & $\omega^2$ & $\omega$  & $\omega^2$ & $-1$ & 1 & $-\omega$  & $-\omega^2$ &  $\omega$  & $\omega^2$ & $-1$ & $-\omega$  & $-\omega^2$ \\ \hline
$\bm{1}_{--2}$ & 1 & 1 & $\omega^2$ & $\omega$  & $\omega^2$ & $\omega$  & $-1$ & 1 & $-\omega^2$ & $-\omega$  &  $\omega^2$ & $\omega$  & $-1$ & $-\omega^2$ & $-\omega$  \\ \hline
$\bm{2}_0$ & 2 & $-2$ & 2 & 2 & $-2$ & $-2$ & 0 & 0 & 0 & 0 & 0 & 0 & 0 & 0 & 0 \\ \hline
$\bm{2}_1$ & 2 & $-2$ & $2\omega$  & $2\omega^2$ & $-2\omega$  & $-2\omega^2$ & 0 & 0 & 0 & 0 & 0 & 0 & 0 & 0 & 0  \\ \hline
$\bm{2}_2$ & 2 & $-2$ & $2\omega^2$ & $2\omega$  & $-2\omega^2$ & $-2\omega$  & 0 & 0 & 0 & 0 & 0 & 0 & 0 & 0 & 0   \\\hline\hline
\end{tabular}}
\end{center}
\caption{Character table of $D_4\times Z_3$.}
\label{tab:cha-tab-D4xZ3-FP4}
\end{table}%

\noindent
The Kronecker products of different representations are as follows:
\begin{align}
\nonumber&\bm{1}_{mnp}\otimes\bm{1}_{m'n'p'}=\bm{1}_{(mm')(nn')[p+p']}\,,~~~\bm{1}_{mnp}\otimes \bm{2}_{p'}=\bm{2}_{[p+p']}\,,\\
&\bm{2}_{p}\otimes \bm{2}_{p'}= \bm{1}_{++[p+p']}\oplus \bm{1}_{+-[p+p']}\oplus \bm{1}_{-+[p+p']}\oplus \bm{1}_{--[p+p']}\,.
\end{align}
where the integer $[n]\equiv n~(\text{mod}~3)$. The tensor products between singlets and doublets are given by
\begin{align}
\nn
\bm{1}_{++p}\otimes \bm{2}_{p'}=\bm{2}_{[p+p']}=
\alpha\begin{pmatrix}
\beta_1\\
\beta_2
\end{pmatrix}\,,
&&\bm{1}_{+-p}\otimes \bm{2}_{p'}=\bm{2}_{[p+p']}=
\alpha\begin{pmatrix}
\beta_1\\
-\beta_2
\end{pmatrix}\,, \\
\bm{1}_{-+p}\otimes \bm{2}_{p'}=\bm{2}_{[p+p']}=
\alpha\begin{pmatrix}
\beta_2\\
\beta_1
\end{pmatrix}\,,
&&\bm{1}_{--p}\otimes \bm{2}_{p'}=\bm{2}_{[p+p']}=
\alpha\begin{pmatrix}
\beta_2\\
-\beta_1
\end{pmatrix}\,.
\end{align}
The tensor products between the doublets are given by
\begin{align}
\bm{2}_{p}\otimes \bm{2}_{p'}= \bm{1}_{++[p+p']}\oplus \bm{1}_{+-[p+p']}\oplus \bm{1}_{-+[p+p']}\oplus \bm{1}_{--[p+p']},~~~~\left\{
\begin{array}{c}
\bm{1}_{++[p+p']}=~\alpha_1\beta_2+\alpha_2\beta_1  \\
\bm{1}_{+-[p+p']}=~\alpha_1\beta_2-\alpha_2\beta_1 \\
\bm{1}_{-+[p+p']}=~\alpha_1\beta_1+\alpha_2\beta_2\\
\bm{1}_{--[p+p']}=~\alpha_1\beta_1-\alpha_2\beta_2
\end{array}\right.\,.
\end{align}
An equivalent basis $(a',b',c')$ for the $D_4\times Z_3$ generators is obtained by the similarity transformation  acting on doublets:
\be
a'=U^{-1} a~ U=
\left(\begin{array}{cc}
0 & -1 \\
1 & 0 \end{array}\right)
~~~~~~~b'=U^{-1} b~ U=
\left(\begin{array}{cc}
-1 & 0 \\
0 & 1 \end{array}\right)~~~~~~~c'=U^{-1} c~ U=c,
\ee
where $U$ is the matrix
\be
U=\frac{1}{\sqrt{2}}
\left(
\begin{array}{cc}
-i&1\\
i&1
\end{array}
\right).
\ee
Singlet representations are unchanged in this new basis, where the consistency conditions of CP transformation are solved by $X_I'=\mathbbm{1}$, both for singlets and doublets.

\newpage
\setcounter{equation}{0}
%%%%%%%%%%%%%%%%%%%%%%%%%%%%%%%%%%%%%%%%%%%%%%%%%%%%%%%%%%%
\section{\label{check} Lepton model of ref.~\cite{Ding:2021iqp}}
\label{amodel}
%%%%%%%%%%%%%%%%%%%%%%%%%%%%%%%%%%%%%%%%%%%%%%%%%%%%%%%%%%%
To verify the above general analysis, we take the lepton model in ref.~\cite{Ding:2021iqp} as an example,
\begin{align}
\nonumber&\rho_{E^c}=\mathbf{2}\oplus\mathbf{1},~~~ \rho_{L} = \mathbf{3'},~~~\rho_{H_u}=\rho_{H_d}=\mathbf{1}\,,\\
&k_{H_u}=k_{H_d}=0,~~~ k_{E_D^c}=-3,~~k_{E_3^c}=k_{L}=-1\,.
\end{align}
In this model, the superpotential $w$ are required to be modular invariant, i.e. $\bm{r}_s=\bm{1}$. The weighted representation matrices can be obtained from the appendix of ref~\cite{Ding:2021iqp} , which happen to have the following diagonal form:
\begin{equation}
\Omega_L(a)=\begin{pmatrix}
-1 & 0 & 0  \\ 0 & 1 & 0 \\ 0 & 0 & 1
\end{pmatrix}\,,\qquad \Omega_L(b)=\begin{pmatrix}
-1 & 0 & 0  \\ 0 & -1 & 0 \\ 0 & 0 & -1
\end{pmatrix}\,,
\end{equation}
\begin{equation}
\Omega_{E^c}(a)=\begin{pmatrix}
-1 & 0 & 0  \\ 0 & 1 & 0 \\ 0 & 0 & 1
\end{pmatrix}\,,\qquad \Omega_{E^c}(b)=\begin{pmatrix}
-1 & 0 & 0  \\ 0 & -1 & 0 \\ 0 & 0 & -1
\end{pmatrix}\,,
\end{equation}
Namely, they are decomposed under $D_4$ into
\begin{equation}
\Omega_L\sim\mathbf{1}^\prime_{+} \oplus \mathbf{1}_{-} \oplus \mathbf{1}_{-} \,,\qquad
\Omega_{E^c}\sim\mathbf{1}^\prime_{+} \oplus \mathbf{1}_{-} \oplus \mathbf{1}_{-}\,,
\end{equation}
they indeed belong to one of the four cases in eq.~\eqref{eq:4decomposition}. The linearized moduli $u_{1,3}$ are transformed as $ \bm{1}'_- \oplus \bm{1}_+$. As we analyzed in Section~\ref{sec:MassMatrix}, the charged lepton mass matrix $m_e$ and neutrino mass matrix $m_\nu$ have the pattern {\bf A} of eq.~\eqref{mA}:
\begin{align}
\nonumber
m_{\bar ee}(u_1,u_3) &\simeq \begin{pmatrix}
1+u_3+u_3^2+u_1^2 & u_1+u_1u_3 & u_1+u_1u_3  \\ u_1+u_1u_3 & 1+u_3+u_3^2+u_1^2 & 1+u_3+u_3^2+u_1^2 \\ u_1+u_1u_3  & 1+u_3+u_3^2+u_1^2 & 1+u_3+u_3^2+u_1^2
\end{pmatrix} +\mathcal{O}(u_i^3) \\
& \simeq \begin{pmatrix}
1& u_1 & u_1  \\ u_1 & 1 & 1 \\ u_1  & 1 & 1
\end{pmatrix} +\dots\,.
\end{align}
\begin{align}
\nonumber
m_\nu(u_1,u_3) &\simeq \begin{pmatrix}
1+u_3+u_3^2+u_1^2 & u_1+u_1u_3 & u_1+u_1u_3  \\ u_1+u_1u_3 & 1+u_3+u_3^2+u_1^2 & 1+u_3+u_3^2+u_1^2 \\ u_1+u_1u_3  & 1+u_3+u_3^2+u_1^2 & 1+u_3+u_3^2+u_1^2
\end{pmatrix} +\mathcal{O}(u_i^3) \\
& \simeq \begin{pmatrix}
1& u_1 & u_1  \\ u_1 & 1 & 1 \\ u_1  & 1 & 1
\end{pmatrix} +\dots\,.
\end{align}
Here we omit the unknown constant coefficients.
Through the $u$-expansion of the Siegel modular forms:
\begin{align}
\nonumber
&((1-u_1)^2-u_3^2)^{-2}Y^{(2)}_{\mathbf{3},1}(u)=15.68u_1 + \mathcal{O}(u_i^3)\,, \\
\nonumber
&((1-u_1)^2-u_3^2)^{-2}Y^{(2)}_{\mathbf{3},2}(u)=1.21+25.49u_1^2-6.37u_3^2 + \mathcal{O}(u_i^3)  \,,\\
\nonumber
&((1-u_1)^2-u_3^2)^{-2}Y^{(2)}_{\mathbf{3},3}(u)=-2.56+5.54u_3-27.04u_1^2 + \mathcal{O}(u_i^3)  \,,\\
&((1-u_1)^2-u_3^2)^{-2}Y^{(2)}_{\mathbf{1}}(u)=1.81+3.92u_3+19.12u_3^2 + 19.12u_1^2+ \mathcal{O}(u_i^3)  \,.
\end{align}
We can obtain the $u$-expansion form of mass matrices in ref.~\cite{Ding:2021iqp},
\begin{align}
\nonumber
m_e(u_1,u_3)  &\simeq \begin{pmatrix}
2.57 & 71.84 u_1 & -101.60 u_1  \\ -71.84u_1 & 23.12 & 18.17 \\ 0.18u_1  & 0.014 & -0.03
\end{pmatrix} v_d +\dots\,. \\
m_\nu(u_1,u_3) &\simeq \begin{pmatrix}
2.01& 22.17u_1 & 15.68u_1  \\ 22.17u_1 & -1.40 & 1.21 \\ 15.68u_1  & 1.21 & 7.97
\end{pmatrix} \frac{v_u^2}{\Lambda} +\dots\,.
\end{align}
where the free Lagrangian parameters are fixed at their best-fit values.~\footnote{The best-fit values for Lagrangian parameters are $\alpha=1,~\beta=-0.83991,~\gamma=0.01176,~g_1=1,~g_2=1.58030$. And the best-fit values of the moduli vacua are $\tau_1=\tau_2=-0.03376+1.11329i,\tau_3=-0.02376+0.50670i$, i.e. $u_1=u_2=-0.00418 + 0.01298i, u_3=-0.02895+0.00474i$.}
This result is consistent with our above general analysis. Since this pattern is not optimal for reproducing the data, the coefficients multiplying $u_1$ are not of the same order, but appear to be tuned to overcome the bad scaling
shown in table~\ref{synops}.

\end{appendix}

\vfill
\newpage

%\bibliographystyle{utphys}
%\bibliography{references}

\providecommand{\href}[2]{#2}\begingroup\raggedright\endgroup

\end{document}